\documentclass[a4paper, amsfonts, amssymb, amsmath,  showkeys, superscriptaddress, nofootinbib]{revtex4-2}
\usepackage{hyperref}
\usepackage[english]{babel}
\usepackage[utf8]{inputenc}
\usepackage[colorinlistoftodos, color=green!40, prependcaption]{todonotes}
\usepackage{dsfont}
\usepackage{caption}
\usepackage{comment}
\usepackage{amsthm}
\usepackage{mathtools}
\usepackage{physics}
\usepackage{xcolor}
\usepackage{graphicx}
\usepackage[left=23mm,right=13mm,top=35mm,columnsep=15pt]{geometry} 
\usepackage{adjustbox}
\usepackage{placeins}
\usepackage[T1]{fontenc}
\usepackage{lipsum}
\usepackage{csquotes}
\usepackage{natbib}

\makeatother

\begin{document}
\title{Multiferroic kinks and spin-flop transition in Ni$_{2}$InSbO$_6$ from first principles}

\author{Ryota Ono}
    \email[Correspondence email address: ]{ryota.ono.gm@gmail.com} 
    \affiliation{Quantum Materials Theory, Italian Institute of Technology, Via Morego 30, 16163 Genova, Italy}
    \affiliation{Research Center for Materials Nanoarchitectonics (MANA), National Institute for Materials Science (NIMS), 1-1 Namiki, Tsukuba, Ibaraki 305-0044, Japan}
\author{Igor Solovyev}
    \affiliation{Research Center for Materials Nanoarchitectonics (MANA), National Institute for Materials Science (NIMS), 1-1 Namiki, Tsukuba, Ibaraki 305-0044, Japan}
\author{Sergey Artyukhin}
    \affiliation{Quantum Materials Theory, Italian Institute of Technology, Via Morego 30, 16163 Genova, Italy}
\date{\today}

\newpage

\begin{abstract}
Magnetoelectric multiferroics are key materials for next-generation spintronic devices due to their entangled magnetic and ferroelectric properties. Spiral multiferroics possess ferroelectric polarization and are particularly promising for electric control of magnetism and magnetic control of ferroelectricity. In this work, we uncover long-period incommensurate states characterized by unique multiferroic kinks in corundum nickelate Ni$_{2}$InSbO$_6$, a member of a promising family of polar magnets. Utilizing a 2-orbital $S=1$ model, we derive formulas for Heisenberg and anisotropic magnetic exchanges and magnetically-induced polarization, enabling their calculations from first principles. We use these parameters in Monte Carlo and Landau theory-based calculations to reproduce experimentally observed magnetic structures and polarization dependence on the magnetic field. We predict magnetic phase transitions between flat spiral, conical spiral, canted antiferromagnetic and ferromagnetic states  under increasing magnetic fields. Kinks in the spiral phases repel each other through a Yukawa-like potential arising from exchange of massive magnons. We also find that suitably directed electric fields can be used to stabilize the ferromagnetic and spiral states. The findings open a new pathway to predictive first-principles modelling of multiferroics and will inspire experiments and technological applications based on multiferroic kinks.

\end{abstract}
\maketitle

\section{INTRODUCTION}
In magnetoelectric multiferroics, a magnetic order coexists and interacts with a ferroelectric one. Several microscopic scenarios of why such coexistence may occur and how the magnetic order can affect the electric polarization have been established~\cite{Kimura2003,Tokura_ad,Tokura2014,Khomskii2009classifying,Cheong2007} and the work is rapidly progressing in this direction. Understanding such interactions between the magnetic and electric degrees of freedom is of great importance from both the fundamental and practical points of view. A special attention is paid to the mutual control of the magnetic structure and the electric polarization by applying the magnetic or electric field. For instance, the external magnetic field can control the magnetic structure, while also changing the ferroelectric polarization. Conversely, the external electric field can be used to control the magnetic structure~\cite{Fiebig_2016}.

There are two main types of multiferroic (MF) materials \cite{Khomskii2009classifying}. In type-I multiferroics the crystal structure itself is ferroelectric, irrespectively of the magnetism. However, the electric polarization can still be controlled by changing the magnetic structure. In type-II MF the crystal structure is centrosymmetric, but the inversion symmetry can still be broken by a magnetic order, which leads to the ferroelectric polarization. An interesting aspect of the type-I materials is that many of them develop chiral magnetic structure, driven by antisymmetric Dzyaloshinskii-Moriya (DM) interactions in the non-centrosymmetric crystal structure. This chirality can be controlled by the magnetic field, presenting another interesting avenue for magnetoelectric control in type-I materials. For instance, a very special type of the chiral magnetic structure is the skyrmion lattice, which has been intensively studied in the context of MF applications in Cu$_2$OSeO$_3$ and GaV$_4$S$_8$~\cite{Cu2OSeO3,GVS}.

Ni$_{2}$InSbO$_6$ (NISO) is one of such chiral MFs. Its low temperature structure has a polar (non-centrosymmetric)  rhombohedral  $R3$ space group ~\cite{NISO_chem,NISO_crys} similar to well known MF corundum derivative Ni$_{3}$TeO$_6$~\cite{SergeyNTO1,SergeyNTO2}.  A number of polar corundum derivatives have recently been synthesized and materials with above room temperature magnetism have been found\cite{Corundum, Frank2022}. These are promising candidates for magnetoelectric applications, in some of which polarization switching has been predicted \cite{Corundum_Vanderbilt}.
Previous studies~\cite{PRB_NISO} revealed an incommensurate antiferromagnetic proper-screw spiral (helix) within each Ni layer, with a long periodicity of 30 unit cells.
The polarization along the threefold rotation axis changes quadratically with the magnetic field due to variations in the spiral order induced by the field~\cite{PRB_NISO}.
In addition, recent experiments have revealed a spin-flop (SF) transition upon applying the magnetic field along the threefold rotation axis~\cite{PRB_NISO_2, PRB_NISO_3}.
However, the detailed microscopic analysis of these observations is lacking. 
The phenomenological mechanisms of the magneto-electric coupling considered so far~\cite{strict, KNB, Arima} are not universal and are influenced by system-dependent factors.
Therefore, it is important to construct realistic models of these materials with interacting spins and magnetically-induced polarization, starting from the modern theory of polarization in terms of Berry phases and Wannier centers~\cite{King,Vdb,Resta} and using model parameters obtained from first-principles calculations.
According to such calculations in the generalized gradient approximation (GGA), NISO can be regarded as an $S=1$ material. In the corundum structure, each Ni$^{2+}$ ion is located in a distorted octahedron of O$^{2-}$ ions. Therefore, the Ni $3d$ states split into triply degenerate $t_{2g}$ and double degenerate $e_{g}$ manifolds. The $t_{2g}$ states are fully occupied and do not significantly contribute to magnetism. On the other hand, the $e_{g}$ states are half-filled and form a group of narrow bands near the Fermi level, which are mainly responsible for the $S=1$ physics. Thus, we can greatly simplify our analysis by constructing  the 2-orbital model for these bands and extracting all parameters of the model from the first principles calculations in the Wannier basis.

Here we model the exchange interactions \emph{and} magnetically-induced polarization emerging from such realistic 2-orbital $e_{g}$ model at the half-filling. 
After extracting the parameters of electronic Hubbard-like model from the first principles calculations, we employ the superexchange theory, which in our case is formulated as a first-order perturbation theory for the Wannier functions with respect to the hopping parameters. For the exchange interactions, the treatment is equivalent to the standard second-order perturbation theory for the magnetic energy, which in the Wannier basis results in the expression for the spin dependent electric polarization. 
Our models highlight the emergence of intriguing cross-coupling phenomena in NISO. 
Specifically, we explore the SF transitions and a crossover to the multiferroic kink array state, both induced by the external magnetic field along the threefold axis. 
The kinks contribute ferroelectric polarization opposite to that of the collinear state and their energetics can be rationalized in terms of repulsion through the Yukawa-like potential and the competition between the DM and magnetic field setting their chemical potential.
Additionally, using a continuous theory, we explore the possibility of cross-control of magnetic (electric) order by an electric (magnetic) field. 


\section{RESULTS}
\subsection{Basic electronic structure, electronic and spin model for NISO}
Results of electronic structure calculations using the experimental crystal structure, shown in Fig.~1\textbf{a}~\cite{NISO_crys}, in GGA~\cite{PBE,PBE2} with the spin-orbit coupling (SOC) for a nonmagnetic state are illustrated in Fig.~1\textbf{e}. 
These calculations clearly reveal two groups of the Ni $3d$ bands: six $t_{2g}$ bands (per two Ni sites) around $-1$ eV and four $e_{g}$ bands around the Fermi level. 
In NISO, the $t_{2g}$ bands are fully occupied and nonmagnetic, while the magnetic properties are mainly associated with the $e_{g}$ bands.
Therefore, we pick this group of states to construct a realistic Hubbard-type model, which would capture the magnetic behavior of NISO. Such Hubbard model has the form:
\begin{align}
 \label{eqn:low_ene}
    &\hat{\mathcal{H}} = \sum_{ij} \sum_{ab} \sum_{\sigma \sigma'} t_{ij}^{ab \sigma \sigma'} \hat{c}_{i a \sigma} ^{\dagger} \hat{c}_{j b \sigma'} ^{\phantom{\dagger}} + H_{\mathrm{on-site}}, 
\end{align}
where $a$ and $b$ ($=$ 1 or 2) label the $e_g$ orbitals, and $ H_{\mathrm{on-site}}$ stands for the on-site Coulomb interactions, which are specified by the intra-orbital interaction $U$, Hund's coupling $J_H$ and inter-orbital interaction $U' = U-2J_H$~\cite{Kanamori}. 
$\hat{c}_{i a \sigma} ^{\dagger}$ ($\hat{c}_{i a \sigma}$) in Eq.~(\ref{eqn:low_ene}) stands for the creation (annihilation) of an electron on the Wannier orbital $a$ of the Ni site $i$ with the spin $\sigma$. 
The parameters of the one-electron part, $t_{ij}^{ab \sigma \sigma'}$, are defined as the matrix elements of GGA Hamiltonian in the Wannier basis. 
Since the basis is complete for the $e_{g}$ bands, these parameters perfectly reproduce the original band structure in GGA (see Fig.~1\textbf{e}). 
The main sources of the SOC in NISO are the $5p$ states of the heavy In/Sb atoms. 
Therefore, it is important to include the SOC \emph{before} the wannierization, at the level of regular GGA calculations. 
Then, although the Wannier functions are formally associated with the Ni $e_{g}$ states, the SOC of the heavy In/Sb atoms will still contribute to the matrix elements $t_{ij}^{ab \sigma \sigma'}$, which can be diagonal as well as off-diagonal with respect to the spin indices. 

Next, we map this low energy electronic model onto the spin model employing for these purposes the superexchange theory. 
In the atomic limit, the ground state corresponding to the high-spin $S=1$ state at half-filling is described by the single Slater determinant. 
The same holds for the one-electron and one-hole $e_{g}$ states emerging in the superexchange theory in the process of virtual excitations. 
Therefore, here we essentially deal with the one-electron theory. This results in the spin Hamiltoninan:
\begin{align}
 \label{eqn:H_s}
    \mathcal{H}_s = &\sum_{<ij>} \left( -J_{ij} \vec{e}_i \cdot \vec{e}_j + \vec{D}_{ij} \cdot (\vec{e}_i \times \vec{e}_j) + \vec{e}_i \cdot \stackrel{\leftrightarrow}{\Gamma}_{ij} \vec{e}_j \right),
\end{align}
where $\vec{e}_{i}$ stands for the classical spin vector at site $i$, the first term is the isotropic interaction, the second term is the anisotropic Dzyaloshinskii-Moriya interaction and the last term is the (traceless) symmetric exchange anisotropy~\cite{Moriya}.

Similarly, the expression for the magnetically-induced polarization is given by
\begin{align}
 \label{eqn:P_s}
    \vec{P}_s = &\sum_{<ij>} \left( \vec{\mathrm{P}}_{ij} \vec{e}_i \cdot \vec{e}_j +  \stackrel{\leftrightarrow}{\mathbf{\mathcal{P}}}_{ij}\cdot (\vec{e}_i \times \vec{e}_j) + \vec{e}_i \cdot \stackrel{\leftrightarrow}{\mathbf{\Pi}}_{ij} \vec{e}_j \right),
\end{align}
 where the first term describes isotropic exchange striction, while the following terms originate form the antisymmetric and the (traceless) symmetric anisotropy ~\cite{PRB_GVS,PRB_RIS,PRL_IRS}.
 This model is derived in the framework of the modern theory of polarization in solids~
 \cite{King,Vdb,Resta}, using perturbation expansion of the Wannier functions. 
 
Since the $R3$ group has only one three-fold rotation axis (along $z$ in the chosen coordinates), all bonds with the same distance should be transformed into each other by the $\hat{C}_3^z$ rotation. 
To illustrate this, consider the bonds surrounding a Ni ion labeled as $0$ in Fig.~1\textbf{c}.
We index its Ni neighbors in the layer above as $j=1, 2, 3$, which are classified as bond type $\alpha=1$. Conversely, the Ni ions in the layer below, indexed as $j=4,5,6$ and categorized as bond type $\alpha=2$, are at a slightly longer distance.
The vectors connecting ion $0$ with ions $j=1, 2, 3$ are expressed as $\vec{\epsilon}_{0j} = (\epsilon_{0j}^{\parallel} \cos(2\pi j/3), \epsilon_{0j}^{\parallel} \sin(2\pi j/3), \epsilon_{0j}^{\perp})$, where $\epsilon_{0j}^{\parallel}$ and $\epsilon_{0j}^{\perp}$ are the lengths of the vector components parallel and perpendicular to the Ni layer, respectively (see Fig.~1\textbf{d}). 
For $j=4, 5, 6$, similar formula applies, with the same $\epsilon_{0j}^{\parallel}$, but a slightly different $\epsilon_{0j}^{\perp}$ and with the arguments of $\sin$ and $\cos$ incremented by $\pi$.
Then, the DM vectors are as follows:
\begin{align}
    &\vec{D}_{0j}^{\alpha} = \left(d^{\parallel}_{\alpha} \cos\theta'_{j,\alpha}, d^{\parallel}_{\alpha}\sin\theta'_{j,\alpha}, d^{\perp}_{\alpha} \right),
\end{align}
where $\theta'_{j,\alpha}=2\pi j/3  + \theta_\alpha$, $d^{\perp}_\alpha$ and $d^{\parallel}_\alpha$ are bond-dependent parameters. 
The DM vectors are antisymmetric: $\vec{D}_{j0}^{\alpha} = -\vec{D}_{0j}^{\alpha}$. 
Similarly, contributions from the isotropic (Heisenberg) exchange to the magnetically-induced polarization (first term of Eq.~(\ref{eqn:P_s})) are as follows:
\begin{align}
    &\vec{\mathrm{P}}_{0j}^{\alpha} = \left(p^{\parallel}_{\alpha} \cos\theta'_{j,\alpha}, p^{\parallel}_{\alpha}\sin\theta'_{j,\alpha}, p^{\perp}_{\alpha} \right),
\end{align}
with symmetric $\vec{\mathrm{P}}_{j0}^{\alpha}$: $\vec{\mathrm{P}}_{j0}^{\alpha}=\vec{\mathrm{P}}_{0j}^{\alpha}$.

An additional single-ion term is allowed in systems $S>1/2$. 
However, in the subsequent discussion, we neglect this term, expecting its effect to be minor since the orbital angular momentum matrix elements vanish between $e_g$ orbitals, making SOC inactive.
Indeed, in our 2-orbital model, the energy splitting due to SOC inside $S=1$ triplet results in very tiny $\Delta E \approx 7$~$\mu$eV (see Supplementary Note 2). 
Turning to the magnetically-induced polarization, we calculate the single-ion term by the formula given in Ref.~\cite{Igor_BCGO}. 
However, we also neglect this term as it is spin-independent in this 2-orbital model.

We derive analytical formulas for the symmetric and antisymmetric exchange constants in (\ref{eqn:H_s}) and magnetically-induced electronic component of the polarization in (\ref{eqn:P_s}) from the 2-orbital model following the strategy used in Refs.~\cite{PRB_RIS, PRL_IRS} (see Methods for the derivation). For exchange interactions we have:
\begin{align}
 \label{eqn:Hs}
    J_{ij} = &\frac{1}{3 \Delta E}  \sum_{\alpha \beta}^{1, 2} \left\{ - 3 \left( t^{0\alpha \beta}_{ij}  \right) ^2 + \mathrm{Tr} (\mathbf{t}^{\alpha \beta}_{ij} \otimes \mathbf{t}^{\alpha \beta}_{ij} ) 
    \right\}, \nonumber \\
\vec{D}_{ij} = &\frac{2}{ \Delta E} \sum_{\alpha,\beta}^{1, 2}  t_{ij}^{0\alpha \beta} \mathbf{t}_{ij}^{\alpha \beta} , \nonumber \\ 
\stackrel{\leftrightarrow}{\Gamma}_{ij} = & \frac{2}{\Delta E}  \sum_{\alpha,\beta}^{1, 2} \left[ \mathbf{t}_{ij}^{\alpha \beta} \otimes \mathbf{t}_{ij}^{\alpha \beta}  - \frac{1}{3}\mathrm{Tr} (\mathbf{t}_{ij}^{\alpha \beta} \otimes \mathbf{t}_{ij}^{\alpha \beta} ) \stackrel{\leftrightarrow}{\mathds{1}} \right],
\end{align}
while the parameters for magnetically-induced electronic component of the polarization have the form:
\begin{align}
\label{eqn:Ps}
\vec{\mathrm{P}}_{ij}  &= -\frac{2e}{3 V \Delta E} \sum_{\alpha, \beta}^{1, 2} \left(3 \vec{r}_{ij}^{0\alpha \beta} t^{0\alpha \beta}_{ij} - \mathrm{Tr}\left( \vec{\mathbf{r}}_{ij}^{ \alpha \beta} \otimes \mathbf{t}_{ij}^{\alpha \beta} \right) \right), \nonumber \\
\vec{\mathbf{\mathcal{P}}}_{ij} &= -\frac{2e}{V \Delta E} \sum_{\alpha, \beta}^{1, 2} \left( t^{0\alpha \beta}_{ij} \vec{\mathbf{r}}_{ij}^{ \alpha \beta}  + \vec{r}_{ij}^{0\alpha \beta} \mathbf{t}_{ij}^{\alpha \beta} \right), \nonumber \\
\stackrel{\leftrightarrow}{\mathbf{\Pi}}_{ij} &= -\frac{2e}{V \Delta E} \sum_{\alpha, \beta}^{1, 2} \left( \vec{\mathbf{r}}_{ij}^{ \alpha \beta} \otimes \mathbf{t}_{ij}^{\alpha \beta} +  
 \mathbf{t}_{ij}^{\alpha \beta} \otimes \vec{\mathbf{r}}_{ij}^{ \alpha \beta} - \frac{2}{3} \mathrm{Tr}\left( \vec{\mathbf{r}}_{ij}^{ \alpha \beta} \otimes \mathbf{t}_{ij}^{\alpha \beta} \right) \stackrel{\leftrightarrow}{\mathds{1}} \right),
 \end{align}
 where $\Delta E =U+J_H$, $\mathbf{t}_{ij}^{\alpha \beta}$ and $\vec{\mathbf{r}}_{ij}^{\alpha \beta}$ are the spin dependent matrix elements of the hopping and position operator, correspondingly, expanded in Pauli matrices as $\hat{t}_{ij}=  \hat{t}^{0}_{ij} \otimes \hat{\sigma}_0 + \sum_{\gamma}^{x, y, z} i \hat{t}^{\gamma}_{ij}  \otimes \hat{\sigma}_{\gamma}$, $\hat{\vec{r}}_{ij}=  \hat{\vec{R}}^{0}_{ij} \otimes \sigma_0 + \sum_{\gamma'} i \hat{\vec{R}}^{\gamma'}_{ij}  \otimes \hat{\sigma}_{\gamma'}$,
 $\stackrel{\leftrightarrow}{\mathds{1}}$ is a $3 \times 3$ identity matrix. 
$J_{ij}$ and $\vec{\mathrm{P}}_{ij}$ are the symmetric isotropic (Heisenberg) exchange (scalar) and exchange striction (vector), while the consecutive terms describe antisymmetric (Dzyaloshinskii-Moriya) exchange and symmetric anisotropic (Ising-like) interactions, respectively.
The exchange parameters obtained with this method are summarized in Tables~\ref{tab:param} and~\ref{tab:param_pol}.
We find that the significant exchange interactions in NISO predominantly originate from the bonds between Ni2 and its Ni neighbors in the layer above (ions 1,2,3, exchange constant $J_1$) and below (ions 4,5,6, exchange constant $J_2$), as shown in Fig.~1\textbf{c}. 
The obtained parameters obey $J_2 < J_1$, which can be understood considering geometric Ni-O-Ni angles, $\angle$(Ni-O-Ni, $J_1$)~$= 129.38^\circ$ and $\angle$(Ni-O-Ni, $J_2$)~$= 136.75^\circ$. 
Typically, the half-filled 2-orbital model predominantly yields AFM interactions, with FM contributions manifesting as effective suppressions in the AFM iteraction constants.
According to the Goodenough-Kanamori rule~\cite{GK}, the bond angle $\angle$(Ni-O-Ni, $J_1$) being close to $90^\circ$ compared to $\angle$(Ni-O-Ni, $J_2$), leads to $J_1$ receiving compensations by the ferromagnetic contributions. 
Consequently, this results in a smaller magnitude of $J_1$ compared to $J_2$.

We note that the 2-orbital model parameters also provide insight into importance of different terms in Eq.~(\ref{eqn:H_s}) and Eq.~(\ref{eqn:P_s}). 
As for the magnetic energy, isotropic and antisymmetric exchange contributions are found to be non-negligible (see Supplementary Note 1 for the values of the symmetric anisotropy tensor components). 
Actually, the relative magnitudes of isotropic and DM interactions indicate that the symmetric anisotropic exchange is inherently small.
This fact aligns with Moriya's paper \cite{Moriya} and our analytical formulas (\ref{eqn:Hs}). 
These state that the DM interaction is first-order in SOC, whereas the symmetric anisotropy is second-order in SOC.
Among the contributions to the polarization, only the isotropic term is found non-negligible. Therefore, in the following we focus mainly on these terms. The actual values of the anisotropic terms of the magnetically-induced polarization and their small effects are detailed in Supplementary Note 3. 
Consequently, in the following we primarily focus on the isotropic exchange, the DM exchange and the isotropic term of the magnetically-induced polarization.

We note that the antiferromagnetic contribution overestimates the exchange parameters as the 2-orbital model does not take into account ferromagnetic contributions from $t_{2g}$ orbitals and Hund's couplings on non-magnetic ions accurately.
Additionally, the relatively small value of the denominator in superexchange theory, specifically $\Delta E = 2.9$~eV, further contributes to potential overestimations in the overall exchange parameters.
Using these parameters in the Monte-Carlo simulations in a $30 \times 30 \times 1$ simulation box with periodic boundary conditions, we obtain the transition temperature, overestimating the experimental one by approximately a factor of two. 
The transition temperature was identified by the peak in the heat capacity in Monte-Carlo simulations using the calculated parameters (Supplementary Note 5).
Nevertheless, we emphasize that the main physics discussed in the following is purely originating from the relative strength between isotropic and anisotropic exchanges. 
%
%
%
%
%
%
%
%
%

\begin{table}
\captionof{table}{Values of isotropic exchanges  and DM parameters in NISO [meV] 
}\label{tab:param}
\begin{ruledtabular}
\begin{tabular}{cccccc}
Bond $\alpha$ & Dist. & $J_\alpha$ & $d^{\parallel}_\alpha$ & $d^{\perp}_\alpha$ & $\theta_\alpha$~[Deg.]\\
\hline
1 & 3.821 & -6.872 & 1.255 &  1.046 & 138 \\
2 & 3.912 & -13.095 & 1.833 & -1.053 & 64  \\  
\end{tabular}
\end{ruledtabular}
\end{table}
\begin{table}
\captionof{table}{Values of isotropic term for polarization and corresponding parameters in NISO [$\mu C /m^2$]
}\label{tab:param_pol}
\begin{ruledtabular}
\begin{tabular}{cccccc}
Bond $\alpha$ & Dist. & $p^{\parallel}_{\alpha}$  & $p^{\perp}_{\alpha}$ & $\theta_\alpha$~[Deg.]\\
\hline
1 & 3.821 & 449 &-138 &  75 \\
2 & 3.912 & 613 & 242 & 180 \\  
\end{tabular}
\end{ruledtabular}
\end{table}

\begin{figure}[htbp]
\centering
\includegraphics[keepaspectratio, scale=0.25]{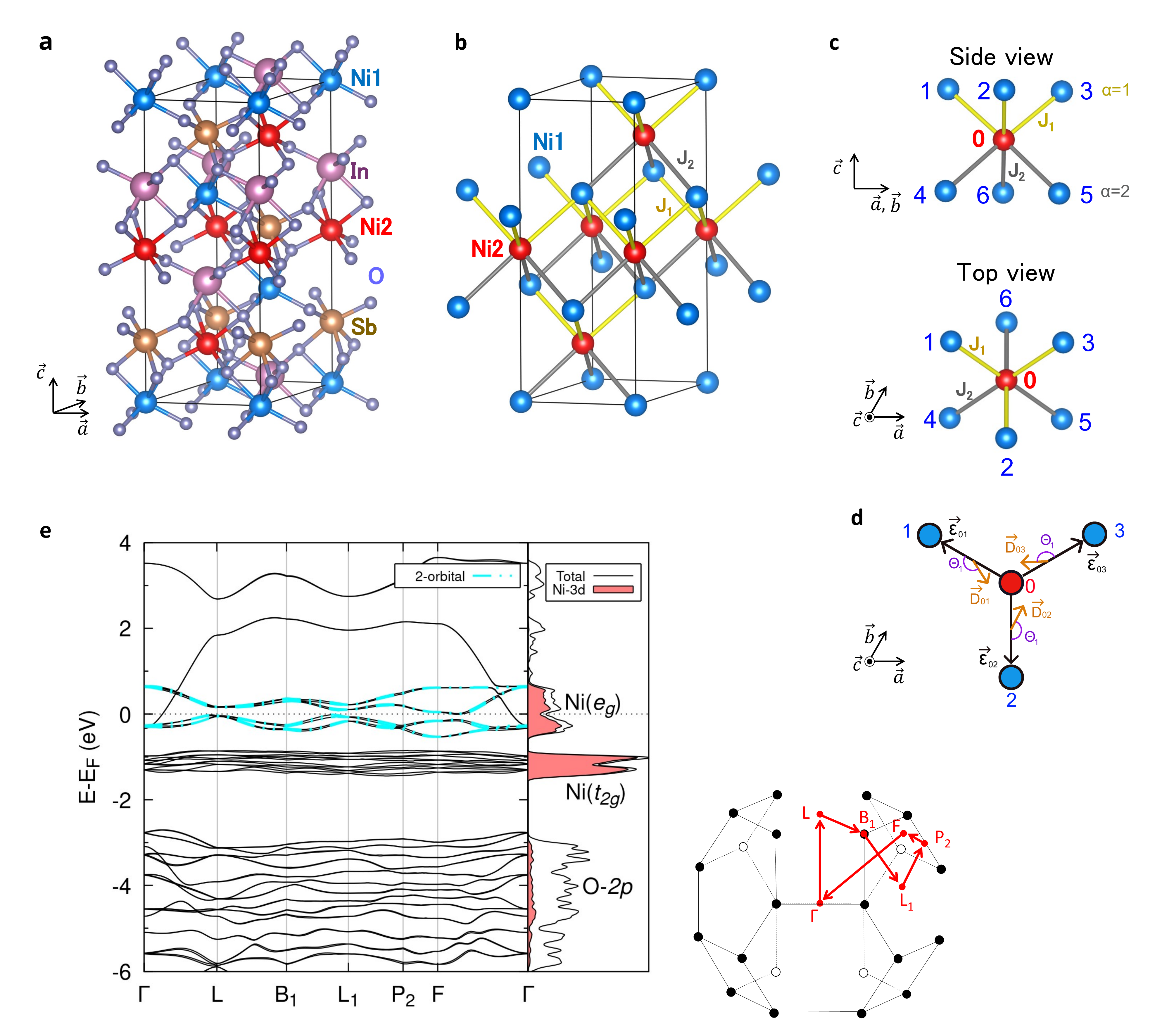}
\end{figure}
\subsection{Magnetic structures in NISO}
We now discuss the magnetic ground state obtained from the single-$q$ spiral Ansatz analysis in NISO.
The 2-orbial parameters show that the $J_1$ and $J_2$ are very strong and AFM. 
Thus, we can expect spins in the neighboring layers to be opposite. 
Since there are four Ni layers in the hexagonal cell, Fig.~1\textbf{a}, the period of the AFM order coincides with the length of the hexagonal c-axis, thus, we have $\vec{q}_{\mathrm{C}} = (0,0,0)$. 
DM interactions modify $q$-vector from a commensurate phase to an incommensurate one by $\delta \vec{q}$. 
The 2-orbital model parameters show that those bonds fulfill a relation $d_{\alpha}^{\parallel} > d_{\alpha}^{\perp}$. 
This stabilizes spin-spiral states with the propagation vector within the layer. 
Namely, the $q$-vector is modified as $\vec{q}_{\mathrm{IC}} = \vec{q}_{\mathrm{C}} + \delta \vec{q} = (\delta q_x, \delta q_y, 0)$.
Then, we can consider the situation given in Fig.~2\textbf{a}, where the spin spiral propagates within a Ni layer.

Magnetic energy contributions from isotropic ($E_{\mathrm{iso}}$) and DM ($E_{\mathrm{DM}}$) interactions are
\begin{align}
\label{eqn:spiralene}
    E_{\mathrm{s}} = &E_{\mathrm{iso}} + E_{\mathrm{DM}} \nonumber \\
    = &\frac{1}{4}\left\{(J_1 + J_2)\left( 12 - \delta q_x^2 - \delta q_y^2 \right) \right. \nonumber  \\ 
    & - 2 \sqrt{3} \delta q_x \left(d^{\parallel}_1  \sin{(\theta_1 - \phi)} +d^{\parallel}_2 \sin{(\theta_2 - \phi)}\right) \nonumber \\
    & \left. -2 \sqrt{3}\delta q_y \left(d^{\parallel}_1  \cos{(\theta_1 - \phi)} +d^{\parallel}_2 \cos{(\theta_2 - \phi)}\right)  \right\}.
\end{align}
Then, we take derivative of the energy with respect to $\delta q_x$ and $\delta q_y$, and find the energy minimum at $\delta \vec{q}$, 
\begin{align}
\begin{dcases}
\label{eqn:dq}
\delta q_x = \sqrt{3} \left(-d^{\parallel}_1  \sin{(\theta_1 - \phi)} -d^{\parallel}_2 \sin{(\theta_2 - \phi)}\right) / (J_1 + J_2)  \\
\delta q_y =  \sqrt{3} \left(-d^{\parallel}_1  \cos{(\theta_1 - \phi)}
-d^{\parallel}_2 \cos{(\theta_2 - \phi)}\right) / (J_1 + J_2).
\end{dcases}
\end{align}

The wavevector components are plotted in Fig.~2\textbf{a} using  the parameters from the 2-orbital model.
We see that, approximately, $(\cos \phi, \sin \phi, 0) \perp \hat{n}^{\perp}$, thus, a cycloidal spiral state (Fig.~2\textbf{b}), while the experiment reported a proper-screw spiral state. 
However, the wave vector and the spiral period, $\delta q_{\mathrm{Ans}} \approx 0.034$ (29 unit cells) are in a good agreement with reported experimental values $\delta q \approx 0.029$ (30 unit cells)~\cite{NISO_chem}. 
Additionally, the symmetric anisotropic interactions tilt the spin-spiral plane.
The calculated parameter actually give a small rotation $\approx \pi/8$ (see Supplementary Note 1).
Another parameter set, calculated from Green's function method \cite{Liechtenstein1987}, gives a proper-screw type spiral with a very long wave length of 142 unit cells  ($\delta q_{\mathrm{GF}} \approx 0.007$; see Supplementary Note 4). 
Since the spin spiral type is not important for the following discussion of magnetic kink generation and magnetically-induced polarization, we use a cycloidal spiral as the ground state of NISO. 
We note that in this mean-field analysis, the energy does not depend on the rotation of the spiral plane $\phi$. This can be straightforwardly confirmed by substituting the analytical formula for the spiral wave vector Eq.~(\ref{eqn:dq}) into the energy Eq.~(\ref{eqn:spiralene}). The resulting formula yields $E_\mathrm{s} = 3 \left( (d^{\parallel}_1)^2 +(d^{\parallel}_2)^2  +2 d^{\parallel}_1 d^{\parallel}_2 \cos (\theta_1-\theta_2)+4(J_1 + J_2)^2\right)/(4(J_1 + J_2))$, thus, independent of $\phi$.
\begin{figure}[htbp]
\centering
\includegraphics[keepaspectratio, scale=1]{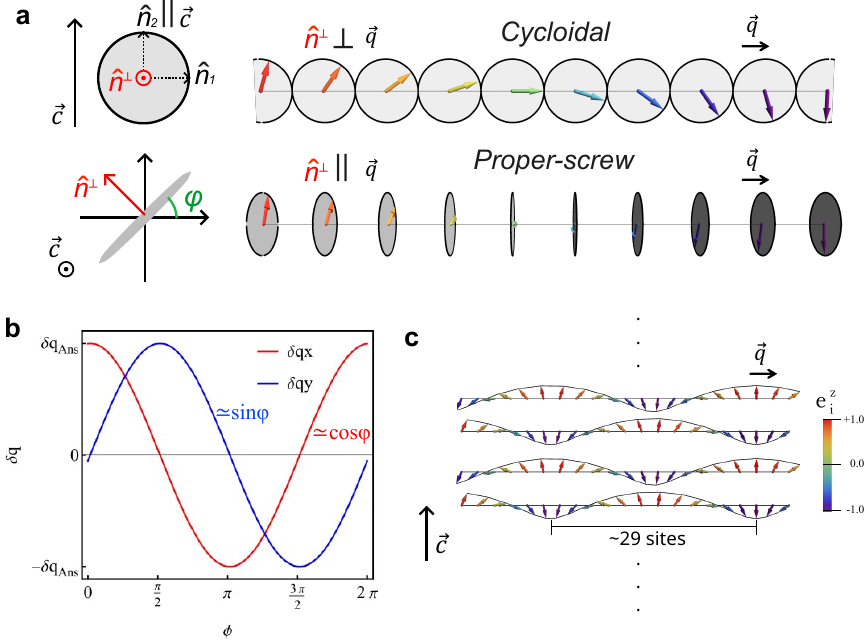}
\label{fig:analytical} 
\end{figure}
%
%
%
%
%
%

In order to understand the magnetic ground state and the response to the external magnetic fields in NISO, we also perform classical Monte-Carlo simulations (MCS) of the spin model Eq.~(\ref{eqn:H_s}) with the 2-orbital parameters. 
Incorporating the Zeeman term $-\sum_i H_z e_i^z$, simulations are carried out on a $30 \times 30 \times 1$ supercell with periodic boundary conditions.

Our MCS reveal the emergence of an anticipated spiral ground state, see Fig.~3\textbf{a}. This result validates the applicability of our single-$q$ spiral Ansatz. 
Upon the application of an external magnetic field perpendicular to the layer, a SF transition is found.
This supports the observations reported in Refs.~\cite{PRB_NISO, PRB_NISO_2, PRB_NISO_3}.
Remarkably, before the SF transition, we observe the solitons, favored by DM interaction, shrinking, while the regions between them, with spins along the magnetic field, experiencing a notable increase (Fig.~3\textbf{b}). 
In comparison to the experimentally determined transition field of approximately 20~T, our calculations yield an overestimated field of approximately 100~T.
This discrepancy is attributed to the overestimated exchange parameters and finite size effects in the MCS.
Despite the success of the MCS, finite size effects significantly influence important spiral deformations by forcing the spiral to be commensurate with the simulation box. 
Therefore, in the following we derive and employ a continuous Ginzburg-Landau-type theory.

\begin{figure}[htbp]
\centering
\includegraphics[keepaspectratio, scale=0.45]{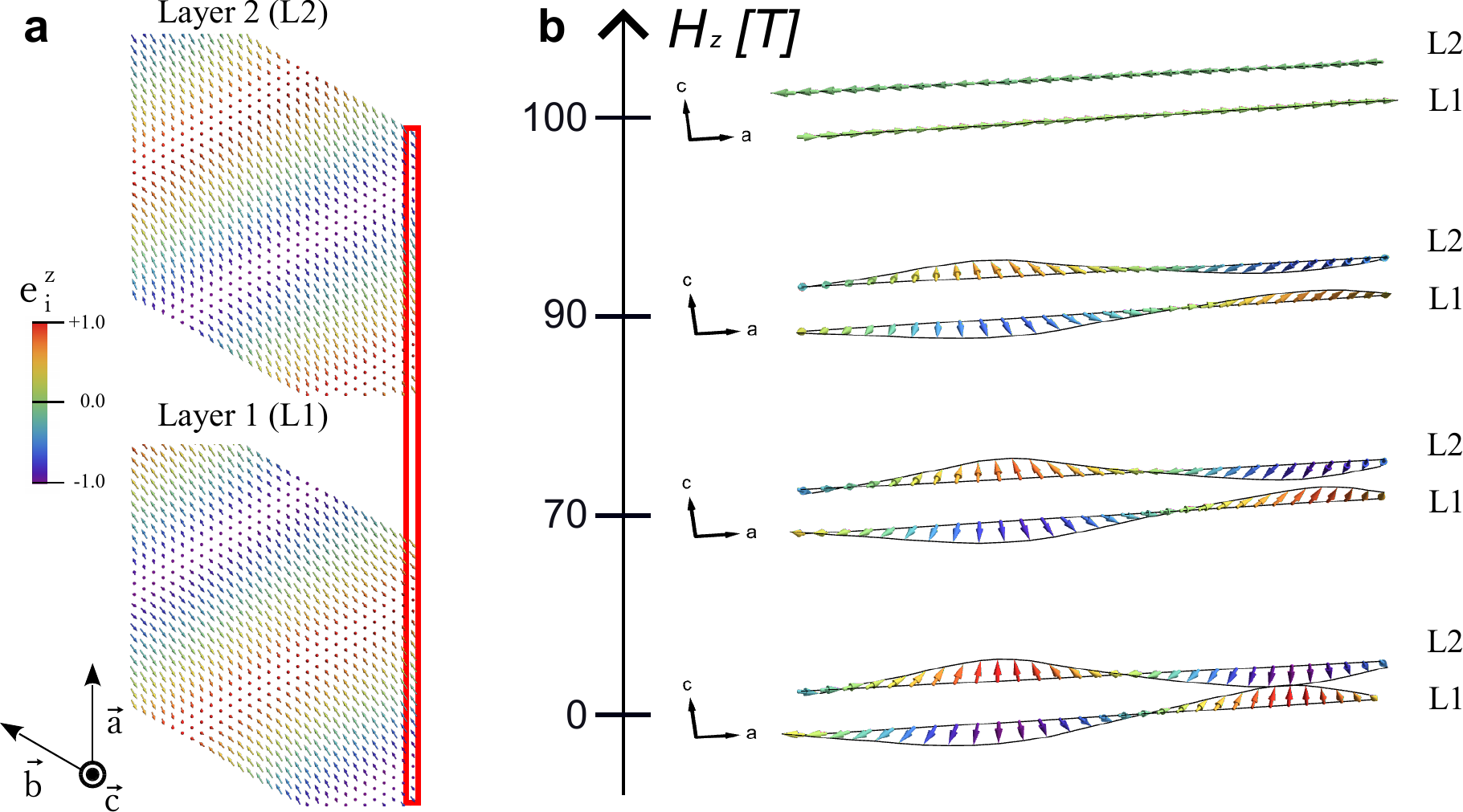}

\end{figure}

\subsection{1D continuous model}
Application of the external magnetic field can modify the spiral period. 
Such a phenomenon, often elusive in MCS due to finite size limitations, can result in overlooking of interesting physics like spiral period changes. 
For instance, De Gennes investigated modifications of the solitonic lattice period by an external magnetic field in liquid crystals~\cite{DeGennes} and concluded that the spiral period follows the analytical expression:
\begin{align}
\label{eqn:DeGennes}
    \frac{L(H_z)}{L(0)} = \left( \frac{2}{\pi} \right)^2 K(k)E(k),
\end{align}
where $L(0)$ is the original period of the perfect spiral at zero field, and $K(k)$, $E(k)$ are elliptic integrals of the first and second kind, respectively.
Here, we formulate a continuous theory to capture such period changes and resulting magnetically-induced polarization without any restriction of the periodicity. 
As we see from MCS, the obtained magnetic structures are quasi-1D, therefore we can effectively model NISO as a 1D-chain. This is due to the following reasons. 
First, the spiral lies in the $xy$-plane.
Second, important exchanges are only $J_1$ and $J_2$ bonds. 
They involve the group of bonds connected by the $\hat{C}_3^z$ operation. 
Those give rise to the energy expression identical to that of a 1D chain.

Fig.~4 shows the spin-spiral ground state and the states it deforms into under an external magnetic field perpendicular to the layer, as obtained by minimizing 1D AFM continuous chain energy in Eq.~(\ref{eqn:Enrec}).
First, application of an external magnetic field generates a kink array state (Fig.~4 \textbf{c, g, k}): instead of rotating in space with the constant wave vector, antiferromagnetic order parameter now rotates in a non-uniform fashion. 
Indeed, when $\vec{A}$ is pointing perpendicular to the magnetic field $\vec{H}$, sublattices cant along the field, which leads to a gain of Zeeman energy, linear in the canting angle (at the expense of the quadratic loss in the antiferromagnetic exchange energy between the sublattices). 
Thus, the regions with  $\vec{A}\perp\vec{H}$ expand, forming plateaus in $A_x$ (Fig.~4 \textbf{c}), which increases the gain of the Zeeman energy on canting. 
In contrast, between the plateaus, kinks in $A_x$ (or solitons) occur where $\vec{A}$ quickly rotates through the field direction. 
At these points, one sublattice aligns with the field and another -- opposite to it, and therefore the gain of Zeeman energy on the canting is not possible. 
The shape of these kinks is analogous to the solitons in nonlinear dynamics because they are solutions of the Sine-Gordon equation. 
The continuous model reveals that with further increase of the field strength, the kinks are pushed apart. At a higher field, the flat spiral turns into a conical one and its plane flops perpendicular to the field (Fig.~4 \textbf{b, f, j}), which enables a higher gain of Zeeman energy from spin canting along the field.
The kinks, present both in the flat and conical spiral phases, can be viewed as particles interacting with each other via exchange of magnons. 
Further increase of the magnetic field strength drives the transition into SF phase.
These transitions are very similar to the one found in our MC simulations (Fig.~3\textbf{b}).
However, in this continuous theory, the transition is much smoother due to unrestricted simulation box size.

\begin{figure}[htbp]
\centering
\includegraphics[keepaspectratio, scale=0.40]{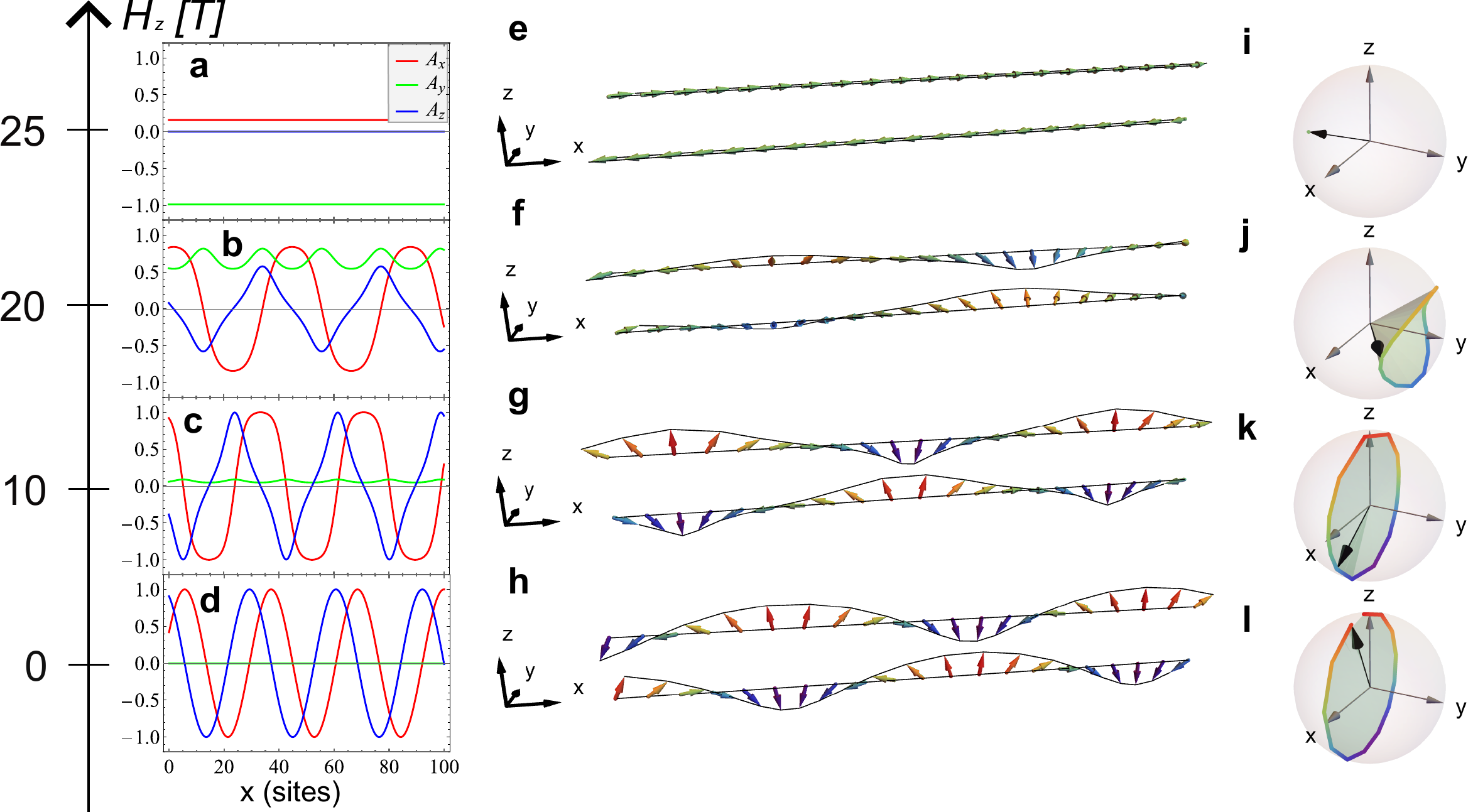}
\end{figure}

In the kink array phase, fast change of the angle $\theta$ near the kink at position $X$ results in a delta function-like change of the gradient $\frac{\partial\theta}{\partial x} \propto \delta (x-X)$. 
In this case, the interactions between kinks are found by solving Euler-Lagrange equation for the magnetic structure deformation, induced by the soliton, and result in the following interaction between kinks,
\begin{align}
\label{eqn:Yukawa}
V(r) = \left( \frac{D}{4 \beta J} \right)e^{-\beta r},
\end{align}
where $\beta = \sqrt{H_z / J} $, and $r$ is the distance between the kinks.
This Yukawa-like interaction can be understood as a long-range repulsion caused by exchanging virtual massive magnons (similar to how Coulomb interaction is caused by exchanges of virtual massless photons). 
High magnetic field results in the enlargement of the area with $\vec{A}\perp \vec{H}$ because of the exponential dependence on $H_z$ in Eq.~(\ref{eqn:Yukawa}).  This is reminiscent of physics found in multiferroic TbFeO$_3$~\cite{Artyukhin2012}.

Fig.~5 illustrates the relationship between spiral period and the polarization change.
Our continuous theory predicts a period change that closely matches De Gennes's analytical formula Eq.~(\ref{eqn:DeGennes}).
Then, the difference in the polarization (computed as a dipole moment of the cluster in Fig.~1\textbf{c} divided by the volume) between spiral and SF states is given as:
\begin{align}
    \Delta P^{\mathrm{iso}}_z = P^{\mathrm{iso}}_{\mathrm{SF},z} - P^{\mathrm{iso}}_{\mathrm{spiral},z} \approx  \pi^2 \delta q^2(p^{\perp}_1 + p^{\perp}_2).
\end{align}
The model parameters result in a very small change due to the isotropic exchange, $\Delta P^{\mathrm{iso}}_z = 0.923$~$\mu C /m^2 $. 
Such a minor polarization change is likely to be imperceptible in experiments.
In the first step (flat spiral phase), a change in polarization is induced by a change in the spiral period (kink distance).
In the second step (conical phase, above 16~T, indicated by a dotted line in Fig.~5), the curve shows a sharp increase of polarization resulting from a ferromagnetic component developing in $xy$-plane due to the SF contribution.

Fig.~6 illustrates the phase diagram in the $(H_z,E_z)$ plane. Here we account for the electric field perturbatively by introducing the lowest order energy density correction $-E_z P_z$.
We find that a small electric field along the ferroelectric polarization stabilizes SF state while an opposite field favors the flat spiral.
Asymmetry with respect to the sign of $E_z$ is caused by the non-centrosymmetric structure of NISO, with the pyroelectric polarization set by a specific cation ordering.
The $P_z$ contributions from the spiral and FM states are opposite, therefore they are stabilized by opposite electric fields.
We note that the large magnetic field continuously deforms the SF state into the FM state. 

\begin{figure}[htbp]
\centering
\includegraphics[keepaspectratio, scale=0.50]{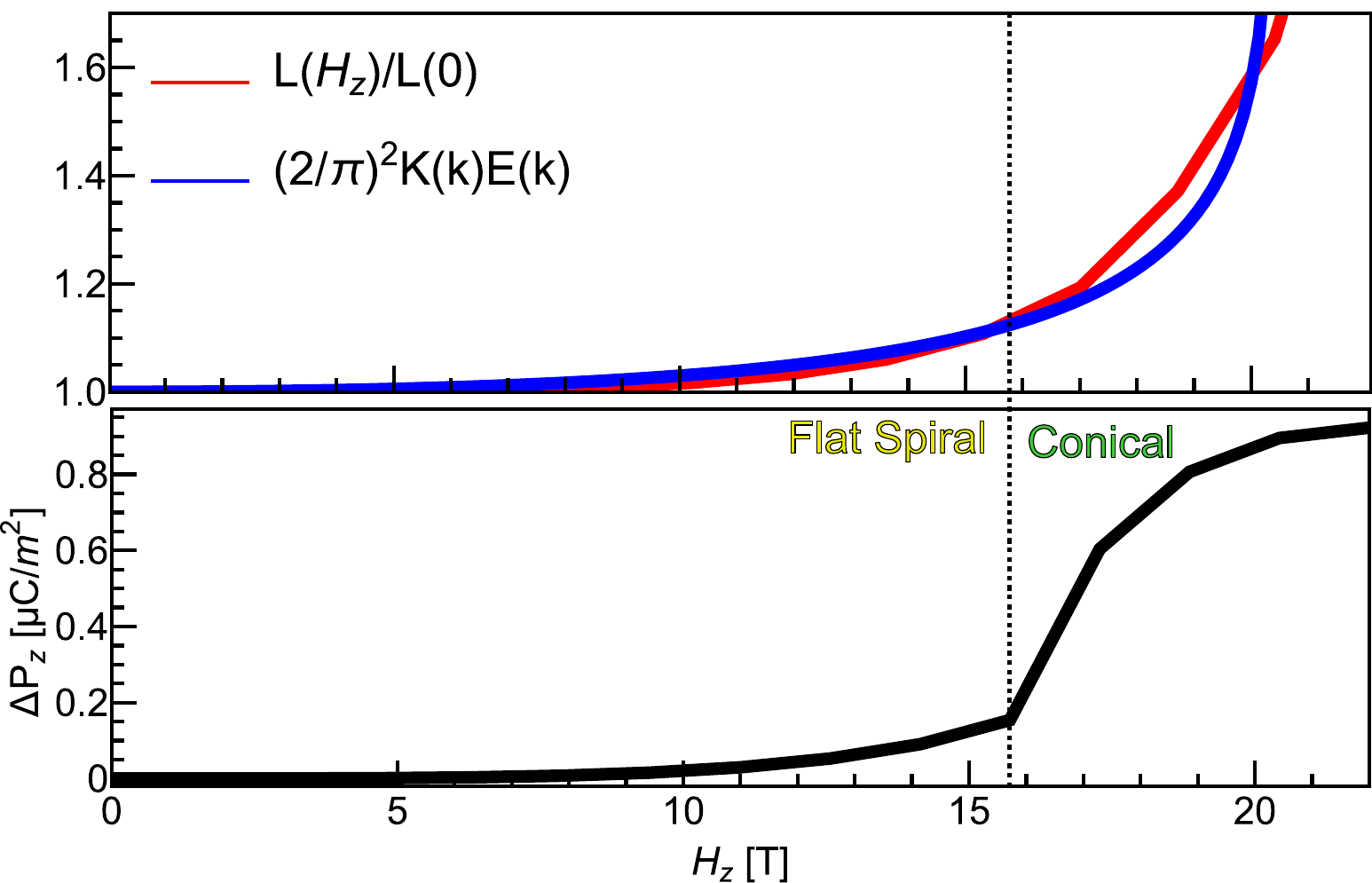}
\end{figure}

\begin{figure}[htbp]
\centering
\includegraphics[keepaspectratio, scale=0.60]{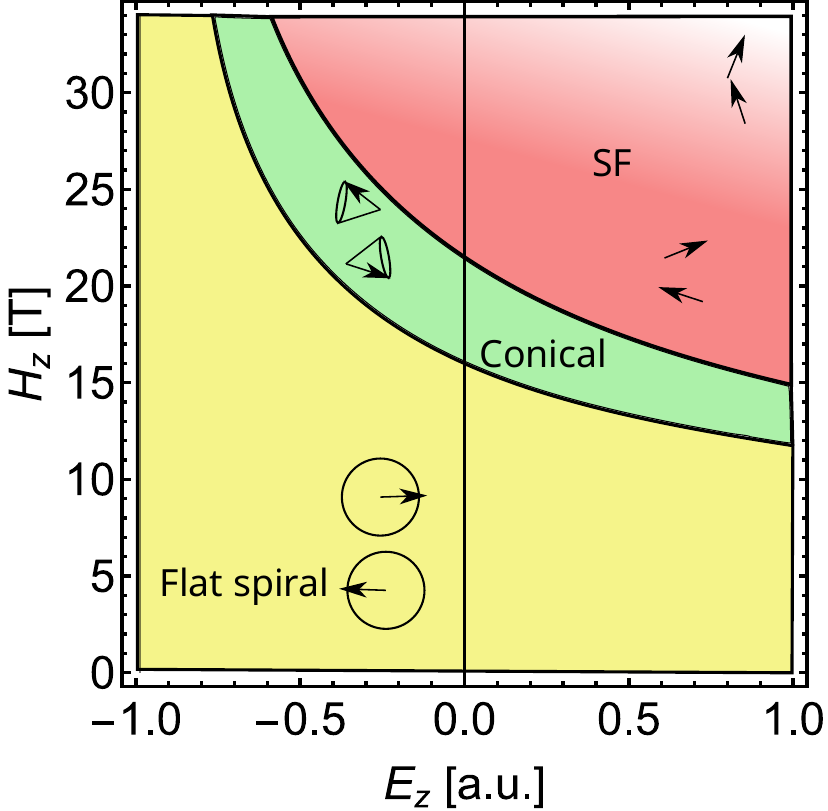}

\end{figure}
%
%
%
%
%
%

\subsection{Polarization change during spin flop-to-FM transition}
Our MCS results indicate the transition from SF phase to the FM state as we apply an external magnetic field (Fig.~7\textbf{a}). 
Notably, the experimental data reveals a much greater polarization change during this transition when compared to the change between the spiral and the SF phases.
During this process, we can express the isotropic exchange contribution to the polarization as follows:
\begin{align}
P^{\mathrm{iso}}_z &= 3 \cos(2\theta) (p^{\perp}_1 + p^{\perp}_2), 
\end{align}
where $\theta$ is the angle between the spin and the $xy$-plane.
In our MCS this angle exhibits a linear dependence on the external magnetic field strength, as depicted in Fig.~7\textbf{b}.
This linear dependence is also found in the experiment~\cite{PRB_NISO}.
Consequently, we anticipate that the magnetically-induced polarization will resemble the behavior shown in Fig.~7\textbf{c}. 

Importantly, the magnitude of this polarization change far exceeds that observed during the spiral $\rightarrow$ SF phase transition.
The polarization change across the sequence of  phase transitions, spiral $\rightarrow$ conical spiral $\rightarrow$ SF $\rightarrow$ FM, reproduces well the experimental data~\cite{PRB_NISO}.

\begin{figure}[htbp]
\centering
\includegraphics[keepaspectratio, scale=0.50]{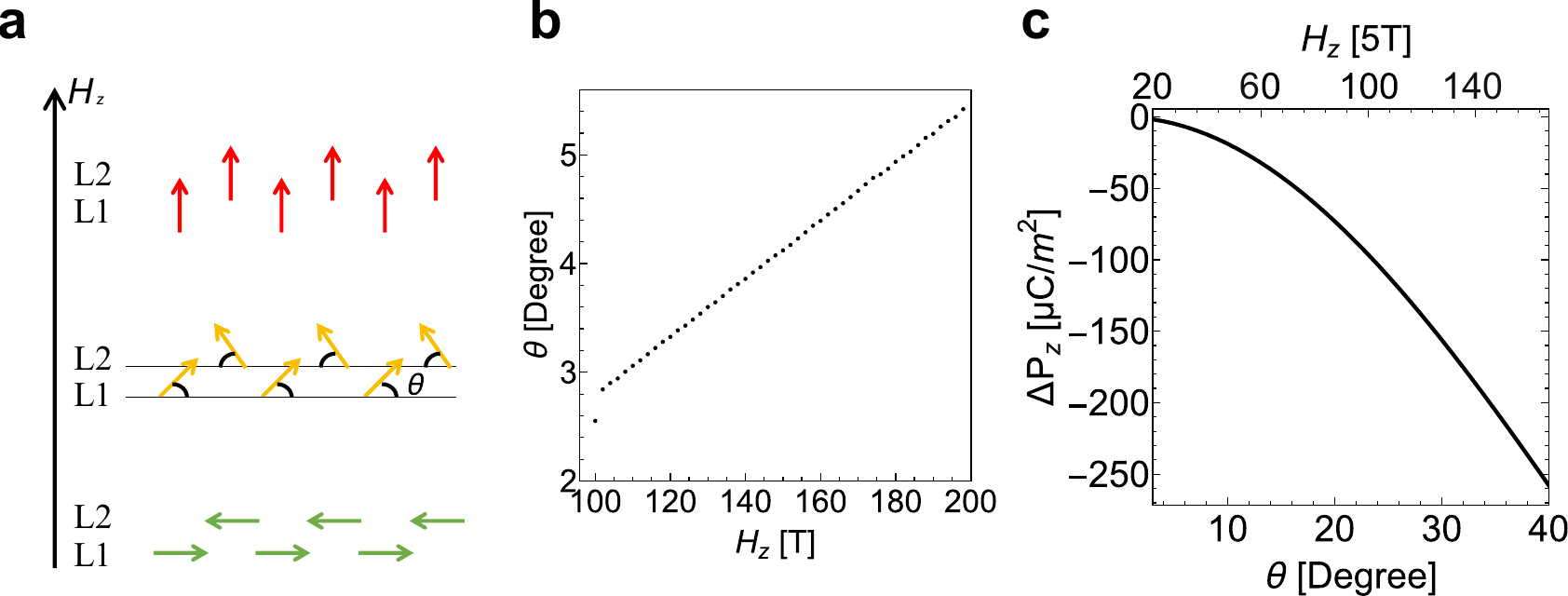}
\end{figure}
\section{Discussion}
Our study of the complex interplay of magnetic and electric fields, magnetic exchanges and DM interactions in corundum nickelate derivative NISO has uncovered intriguing properties of multiferroic kinks and their implications for the field of MFs. 
The methodological advancements allowed a quantitative modelling of key magnetoelectric phenomena in NISO.

We have derived analytical formulas for magnetic exchanges and magnetically-induced polarization, starting from a minimal 2-orbital model, suitable for computing these parameters from first principles. 
These formulas can be applied to other $S = 1$ MFs (e.g. Ni$_3$TeO$_6$,  Ni$_{2}$ScSbO$_6$). 
Additionally, we have developed a continuous theory to address complex changes in the spin-spiral structures, which extends our understanding of spiral MFs. This theory should be applicable to a wide class of materials where spiral state is stabilized by Dzyaloshinskii-Moriya interactions, for example BiFeO$_3$. 
Our analysis starts from the ground state with spiral ordering within the layers.
We find that a spiral structure deforms into an array of multiferroic kinks upon the application of an external magnetic field in the plane of the spiral. 
These kinks play a crucial role in determining the period of the spiral, leading to a small polarization that is opposite to the FM contribution.
Importantly, external magnetic fields can directly control the distances between these kinks.
When a greater magnetic field is applied parallel to the three-fold axis, it induces the SF transition, consistent with experimental observations.

Using analytical and numerical models with parameters calculated from first principles,  we have identified a three-step phase transition (spiral $\rightarrow$ conical spiral $\rightarrow$ SF $\rightarrow$ FM) under an applied magnetic field. 
Our ab-initio model shows that the most important mechanism of the magnetically-induced polarization in NISO is the isotropic exchange striction. 
During the first phase transition, the polarization undergoes a small change, closely related to the magnetic kink distance (spiral period).
The kink density produces a contribution to the polarization that opposes the ferromagnetic contribution, resulting in polarization change similar to the spiral period change.
In the second step, the polarization curve shows a sharp increase related to the FM component development in the $xy$-plane due to SF contribution.
In the third step, the polarization changes significantly, simultaneously with the linear change of the canting angle $\theta$ with the magnetic field. These polarization changes imply the possibility of switching between these magnetic structures by an external electric field, enabling the electric control of magnetism.
The proposed scenario is supported by the excellent agreement with the experimental data on the field dependence of the polarization.

In summary, the work advances the understanding of magnetoelectric effect in spiral multiferroics.
NISO shows a rich phase diagram with flat spiral ground state, magnetic kink arrays, conical kink phase and SF phase. We reveal their connection to the magnetically-induced polarization, and the possibility of electric switching between magnetic phases via an external electric field. 
Given the importance of the manipulation of the spiral MFs by external fields, we expect that our findings will motivate new experiments and facilitate the applications of spiral MFs in the next-generation memory and spintronic devices.

\section{Methods}
\subsection{First principles calculations}
Density functional theory calculations have been performed for the experimental crystal structure~\cite{NISO_crys} (see Fig.~1\textbf{a} for the hexagonal cell) using a rhombohedral cell. 
These calculations employed norm-conserving pseudopotentials within the Quantum ESPRESSO package~\cite{QE1}. 
The plane wave cutoff was set to 1088~eV, the Brillouin zone was sampled by a 7 × 7 × 7 Monkhorst–Pack k-point mesh. 

The hopping parameters of Eq.~(\ref{eqn:low_ene}) is calculated using the Maximally Localized Wannier Fucntion (MLWF) technique~\cite{Marzari} as implemented in the Wannier90 code~\cite{W90,W90_2,Marzari_mod}. 

The screened Coulomb and exchange interactions of Eq.~(\ref{eqn:low_ene}) are calculated using constrained RPA technique~\cite{crpa}, which yields $U = 2.3$~eV and $J_H = 0.6$~eV.

The crystal structures were visualised with VESTA~\cite{VESTA}.

\subsection{Analytical formula for the parameters of spin models}
The second order perturbation energy with respect to electron hopping in the 2-orbital model is formulated as:
\begin{align}
E^{(2)} = -\frac{1}{\Delta E} \sum_{\alpha, \beta}^{1, 2} \left( |\bra{\alpha -,i} \hat{t}_{ij}\ket{\beta +,j}|^2 + |\bra{\alpha -,j} \hat{t}_{ji}\ket{\beta +,i}|^2 \right),
\end{align}
where $\ket{\alpha +(-),i}$ indicates occupied (unoccupied) spin-orbital at site $i$ and $\Delta E = U + J_H$.
Now, we assume Hamiltonian matrix elements $\hat{t}_{ij}$ are ordered as $\ket{1} \ket{2}$ (pairs of Kramers' states). 
Consequently, the corresponding orbital ket vectors are straightforwardly defined as: $ \ket{1} = \begin{pmatrix} 1\\ 0 \end{pmatrix}$ and $ \ket{2} = \begin{pmatrix} 0\\ 1 \end{pmatrix}$ . 
Furthermore, when considering the occupied and unoccupied spin states, we assume them to have general spinor functions:
\begin{align}
\ket{-} = 
\begin{pmatrix}
-\sin \frac{\theta}{2} e^{- i \phi} \\
\cos \frac{\theta}{2}
\end{pmatrix}
\end{align}
and 
\begin{align}
\ket{+} = 
\begin{pmatrix}
\cos \frac{\theta}{2} \\
\sin \frac{\theta}{2} e^{ i \phi}
\end{pmatrix}.
\end{align}
These states have the maximal spin projection in the direction of an associated classical spin $\vec{e} = \bra{+}\hat{\vec{\sigma}}\ket{+} = (\sin\theta \cos\phi, \sin \theta \sin \phi, \cos\theta)$.

Next, we decompose general $4\times4$ hopping matrices (with the proper phase) into their orbital and spin components, as described by the following equations:
\begin{align}
\hat{t}_{ij}=  \hat{T}^{0}_{ij} \otimes \hat{\sigma}_0 + \sum_{\gamma}^{x, y, z} i \hat{T}^{\gamma}_{ij}  \otimes \hat{\sigma}_{\gamma},
\end{align}
with the orbital part
\begin{align}
\hat{T}^{\eta}_{ij} = 
\begin{pmatrix}
t^{\eta 11}_{ij} & t^{\eta 12}_{ij} \\
t^{\eta 21}_{ij} & t^{\eta 22}_{ij}
\end{pmatrix}, (\eta = 0, x, y, z)
\end{align}
where $\hat{\sigma}_0$ is the $2\times 2$ identity matrix and $\hat{\sigma}_\gamma ; \gamma=x,y,z$ are the spin Pauli matrices.
Here, we choose the phases such that non-SOC related terms are represented solely by pure real coefficients, while SOC terms are represented solely by pure imaginary coefficients. 
Then, taking energy difference for different spin orientations and mapping, it is straightforward to find analytical formual Eq.~(\ref{eqn:Hs}).

\indent The spin model can also be derived with respect to spin operators acting on $S=1$ multiplets. 
The procedure is simply finding the correspondence between a general spin Hamiltonian: 
\begin{align}
    \mathcal{H}_{\mathrm{s}} =  \hat{\bm{S}}_i \cdot \stackrel{\leftrightarrow}{\mathds{J}}_{ij} \cdot \hat{\bm{S}}_j,
\end{align}
 and the second order perturbation energy:
\begin{align}
    \mathcal{H}_{\mathrm{eff}} =  -\sum_{jM} \frac{1}{\Delta E_{jM}} \hat{\mathcal{T}}_{ij}\ket{jM}\bra{jM}\hat{\mathcal{T}}_{ji}.
\end{align}
Here, $\stackrel{\leftrightarrow}{\mathds{J}}_{ij}$ is a 3x3 tensor, $\hat{\mathcal{T}}{ij}$ is a hopping integral defined as: $\hat{\mathcal{T}}_{ij} = \sum_{\alpha \beta} \sum_{\sigma \sigma'} t_{ij}^{\alpha \beta \sigma \sigma'} \hat{c}_{i \alpha \sigma} ^{\dagger} \hat{c}_{j \beta \sigma'} $ and $\ket{jM}$ represents the intermediate states $\ket{1/2, m_1} \otimes \ket{1/2, m_2} \otimes \ket{1/2, m_3} \in S=1/2$ at site $j$. 
For instance, $\bra{0,1}\mathcal{H}_{\mathrm{eff}} \ket{1,1} = \frac{1}{\sqrt{2}} (J^{xz}_{ij} +i J^{yz}_{ij})$. 
This approach results in the following expressions:
\begin{align}
J_{ij}^{xx} = &+\frac{1}{2}\frac{1}{ \Delta E}  \sum_{\alpha,\beta}^{1, 2}  \left(+  
t_{ij}^{\alpha \beta \uparrow \downarrow} \left( 
t_{ij}^{\alpha \beta \downarrow \uparrow} \right)^* +
t_{ij}^{\alpha \beta \uparrow \uparrow} \left( t_{ij}^{\alpha \beta \downarrow \downarrow} \right)^* + t_{ij}^{\alpha \beta \downarrow \downarrow} \left( t_{ij}^{\alpha \beta \uparrow \uparrow} \right)^* + t_{ij}^{\alpha \beta \downarrow \uparrow} \left( t_{ij}^{\alpha \beta \uparrow \downarrow} \right)^* \right)\\
J_{ij}^{xy} = &-\frac{i}{2}\frac{1}{ \Delta E}  \sum_{\alpha,\beta}^{1, 2}  \left(-  
t_{ij}^{\alpha \beta \uparrow \downarrow} \left( 
t_{ij}^{\alpha \beta \downarrow \uparrow} \right)^* +
t_{ij}^{\alpha \beta \uparrow \uparrow} \left( t_{ij}^{\alpha \beta \downarrow \downarrow} \right)^* - t_{ij}^{\alpha \beta \downarrow \downarrow} \left( t_{ij}^{\alpha \beta \uparrow \uparrow} \right)^* + t_{ij}^{\alpha \beta \downarrow \uparrow} \left( t_{ij}^{\alpha \beta \uparrow \downarrow} \right)^* \right)\\
J_{ij}^{xz} = &-\frac{1}{2}\frac{1}{ \Delta E}  \sum_{\alpha,\beta}^{1, 2}  \left(+  
t_{ij}^{\alpha \beta \downarrow \downarrow} \left( 
t_{ij}^{\alpha \beta \uparrow \downarrow} \right)^* -
t_{ij}^{\alpha \beta \downarrow \uparrow} \left( t_{ij}^{\alpha \beta \uparrow \uparrow} \right)^* + t_{ij}^{\alpha \beta \uparrow \downarrow} \left( t_{ij}^{\alpha \beta \downarrow \downarrow} \right)^* - t_{ij}^{\alpha \beta \uparrow \uparrow} \left( t_{ij}^{\alpha \beta \downarrow \uparrow} \right)^* \right)\\
J_{ij}^{yx} = &-\frac{i}{2}\frac{1}{ \Delta E}  \sum_{\alpha,\beta}^{1, 2}  \left(-  
t_{ij}^{\alpha \beta \uparrow \downarrow} \left( 
t_{ij}^{\alpha \beta \downarrow \uparrow} \right)^* -
t_{ij}^{\alpha \beta \uparrow \uparrow} \left( t_{ij}^{\alpha \beta \downarrow \downarrow} \right)^* + t_{ij}^{\alpha \beta \downarrow \downarrow} \left( t_{ij}^{\alpha \beta \uparrow \uparrow} \right)^* + t_{ij}^{\alpha \beta \downarrow \uparrow} \left( t_{ij}^{\alpha \beta \uparrow \downarrow} \right)^* \right)\\
J_{ij}^{yy} = &+\frac{1}{2}\frac{1}{ \Delta E}  \sum_{\alpha,\beta}^{1, 2}  \left(-  
t_{ij}^{\alpha \beta \uparrow \downarrow} \left( 
t_{ij}^{\alpha \beta \downarrow \uparrow} \right)^* +
t_{ij}^{\alpha \beta \uparrow \uparrow} \left( t_{ij}^{\alpha \beta \downarrow \downarrow} \right)^* + t_{ij}^{\alpha \beta \downarrow \downarrow} \left( t_{ij}^{\alpha \beta \uparrow \uparrow} \right)^* - t_{ij}^{\alpha \beta \downarrow \uparrow} \left( t_{ij}^{\alpha \beta \uparrow \downarrow} \right)^* \right)\\
J_{ij}^{yz} = &+\frac{i}{2}\frac{1}{ \Delta E}  \sum_{\alpha,\beta}^{1, 2}  \left(+  
t_{ij}^{\alpha \beta \downarrow \downarrow} \left( 
t_{ij}^{\alpha \beta \uparrow \downarrow} \right)^* -
t_{ij}^{\alpha \beta \downarrow \uparrow} \left( t_{ij}^{\alpha \beta \uparrow \uparrow} \right)^* - t_{ij}^{\alpha \beta \uparrow \downarrow} \left( t_{ij}^{\alpha \beta \downarrow \downarrow} \right)^* + t_{ij}^{\alpha \beta \uparrow \uparrow} \left( t_{ij}^{\alpha \beta \downarrow \uparrow} \right)^* \right)\\
J_{ij}^{zx} = &-\frac{1}{2}\frac{1}{ \Delta E}  \sum_{\alpha,\beta}^{1, 2}  \left(-  
t_{ij}^{\alpha \beta \uparrow \uparrow} \left( 
t_{ij}^{\alpha \beta \uparrow \downarrow} \right)^* +
t_{ij}^{\alpha \beta \downarrow \uparrow} \left( t_{ij}^{\alpha \beta \downarrow \downarrow} \right)^* - t_{ij}^{\alpha \beta \uparrow \downarrow} \left( t_{ij}^{\alpha \beta \uparrow \uparrow} \right)^* + t_{ij}^{\alpha \beta \downarrow \downarrow} \left( t_{ij}^{\alpha \beta \downarrow \uparrow} \right)^* \right)\\
J_{ij}^{zy} = &+\frac{i}{2}\frac{1}{ \Delta E}  \sum_{\alpha,\beta}^{1, 2}  \left(-  
t_{ij}^{\alpha \beta \uparrow \uparrow} \left( 
t_{ij}^{\alpha \beta \uparrow \downarrow} \right)^* +
t_{ij}^{\alpha \beta \downarrow \uparrow} \left( t_{ij}^{\alpha \beta \downarrow \downarrow} \right)^* + t_{ij}^{\alpha \beta \uparrow \downarrow} \left( t_{ij}^{\alpha \beta \uparrow \uparrow} \right)^* - t_{ij}^{\alpha \beta \downarrow \downarrow} \left( t_{ij}^{\alpha \beta \downarrow \uparrow} \right)^* \right)\\
J_{ij}^{zz} = &+\frac{1}{2}\frac{1}{ \Delta E}  \sum_{\alpha,\beta}^{1, 2} \left( - 
|t_{ij}^{\alpha \beta \uparrow \downarrow}|^2 - |t_{ij}^{\alpha \beta \downarrow \uparrow}|^2 + |t_{ij}^{\alpha \beta \uparrow \uparrow}|^2 + |t_{ij}^{\alpha \beta \downarrow \downarrow}|^2
\right),
\end{align}
where the difference in signs between the second-order energies arises from particle exchanges.
This tensor can generally be decomposed into isotropic interaction: $J_{ij} = - \frac{1}{3}\mathrm{Tr} \stackrel{\leftrightarrow}{\mathds{J}}_{ij},$ Dzhaloshinskii-Moriya vectors: $
\vec{D}_{ij} = \frac{1}{2} 
\begin{pmatrix}
J_{ij}^{yz} - J_{ij}^{zy} \\
J_{ij}^{zx} - J_{ij}^{xz} \\
J_{ij}^{xy} - J_{ij}^{yx}
\end{pmatrix}
$, and symmetric anisotropy: $\stackrel{\leftrightarrow}{\Gamma}_{ij} =  \frac{1}{2}\left(\stackrel{\leftrightarrow}{\mathds{J}}_{ij} + \stackrel{\leftrightarrow}{\mathds{J}}_{ij}^{\mathrm{t}} \right) - J_{ij} \stackrel{\leftrightarrow}{\mathds{1}}
$. 
However, this approach results in the exactly same result as the classical vector approach Eq.~(\ref{eqn:Hs}).
The equivalence of the two approaches can be easily demonstrated by considering the relationship between the hopping integrals
For example, 
\begin{align*}
 t_{ij}^{\alpha \beta \uparrow \uparrow} =  t^{0 \alpha \beta}_{ij} + i t^{z \alpha \beta}_{ij}.
\end{align*}
Substituting this definition into Eq.~(9-17) yields the same formula as Eq.~(6). Here, we provide an explicit notation for the diagonal elements as an example:
\begin{align*}
J_{ij}^{xx} = &+\frac{1}{2}\frac{1}{ \Delta E}  \sum_{\alpha,\beta}^{1, 2}  \left(+  
t_{ij}^{\alpha \beta \uparrow \downarrow} \left( 
t_{ij}^{\alpha \beta \downarrow \uparrow} \right)^* +
t_{ij}^{\alpha \beta \uparrow \uparrow} \left( t_{ij}^{\alpha \beta \downarrow \downarrow} \right)^* + t_{ij}^{\alpha \beta \downarrow \downarrow} \left( t_{ij}^{\alpha \beta \uparrow \uparrow} \right)^* + t_{ij}^{\alpha \beta \downarrow \uparrow} \left( t_{ij}^{\alpha \beta \uparrow \downarrow} \right)^* \right)\\
=& \frac{1}{ \Delta E}  \sum_{\alpha,\beta}^{1, 2} \left( (t_{ij}^{0 \alpha \beta })^2  + (t_{ij}^{x \alpha \beta })^2 - (t_{ij}^{y \alpha \beta })^2  
- (t_{ij, z}^{\alpha \beta })^2 \right), \\
J_{ij}^{yy} = &+\frac{1}{2}\frac{1}{ \Delta E}  \sum_{\alpha,\beta}^{1, 2}  \left(-  
t_{ij}^{\alpha \beta \uparrow \downarrow} \left( 
t_{ij}^{\alpha \beta \downarrow \uparrow} \right)^* +
t_{ij}^{\alpha \beta \uparrow \uparrow} \left( t_{ij}^{\alpha \beta \downarrow \downarrow} \right)^* + t_{ij}^{\alpha \beta \downarrow \downarrow} \left( t_{ij}^{\alpha \beta \uparrow \uparrow} \right)^* - t_{ij}^{\alpha \beta \downarrow \uparrow} \left( t_{ij}^{\alpha \beta \uparrow \downarrow} \right)^* \right) \\
=& \frac{1}{ \Delta E}  \sum_{\alpha,\beta}^{1, 2} \left( (t_{ij}^{0 \alpha \beta })^2  + (t_{ij}^{y \alpha \beta })^2 - (t_{ij}^{x \alpha \beta })^2  
- (t_{ij, z}^{\alpha \beta })^2 \right), \\
J_{ij}^{zz} = &+\frac{1}{2}\frac{1}{ \Delta E}  \sum_{\alpha,\beta}^{1, 2} \left( - 
|t_{ij}^{\alpha \beta \uparrow \downarrow}|^2 - |t_{ij}^{\alpha \beta \downarrow \uparrow}|^2 + |t_{ij}^{\alpha \beta \uparrow \uparrow}|^2 + |t_{ij}^{\alpha \beta \downarrow \downarrow}|^2
\right)  \\
=& \frac{1}{ \Delta E}  \sum_{\alpha,\beta}^{1, 2} \left( (t_{ij}^{0 \alpha \beta })^2  + (t_{ij}^{z \alpha \beta })^2 - (t_{ij}^{x \alpha \beta })^2  
- (t_{ij}^{y \alpha \beta })^2 \right) .
\end{align*}
Therefore, the isotropic interaction is given as
\begin{align*}
    J_{ij} = - \frac{1}{3}\mathrm{Tr} \stackrel{\leftrightarrow}{\mathds{J}}_{ij} = \frac{1}{3 \Delta E}  \sum_{\alpha \beta}^{1, 2} \left\{ - 3 \left( t^{0\alpha \beta}_{ij}  \right) ^2 + \mathrm{Tr} (\mathbf{t}_{ij}^{\alpha \beta} \otimes \mathbf{t}_{ij}^{\alpha \beta} ) 
    \right\}.
\end{align*}
All the other expression can be restored following the same procedure.

A similar theory can be developed for magnetically induced electronic polarization. 
The main idea is to expand the Wannier functions with respect to first-order of the hopping and apply the general theory of electronic polarization in solids. 
This expansion corresponds to spin-dependent modulations of the Wannier density, resulting in changes in the Wannier centers. 
Starting point of the theory is the general theory for polarization in solids~\cite{Vdb}:
\begin{align}
 \label{eqn:pol}
\vec{P} = -\frac{e}{V} \sum_i^{\mathrm{occ}} \bra{w_i|\hat{r}}\ket{w_i},
\end{align}
where $\bra{w_i|\hat{r}}\ket{w_i}$ represents the position operator's diagonal elements in the Wannier basis, and $V$ is the cell volume.
Next, the first-order perturbation expansion of the Wannier function with respect to electron hopping is:
\begin{align}
 \label{eqn:exp}
\ket{w_i} \approx \ket{\alpha +, i} - \frac{1}{\Delta E} \sum_j \sum_{\beta}^{1, 2}  \ket{\beta -, j} \bra{\beta - , j} \hat{t}_{ji}\ket{\alpha +, i}.
\end{align}
By substituting Eq.~(\ref{eqn:exp}) into Eq.~(\ref{eqn:pol}), we obtain pair and single-ion terms as $\vec{P} = \sum_i \vec{P}_i + \sum_{<ij>} \vec{P}_{ij}$.
The pair interaction terms are as follows:
\begin{align}
\label{eqn:P_peruturb}
\vec{P}_{ij} = \frac{e}{V \Delta E} \sum_{\alpha \beta}^{1, 2} & \left[ \bra{\alpha +, i}\hat{\vec{r}}_{ij}\ket{\beta -, j} \bra{\beta -, j}\hat{t}_{ji}\ket{\alpha +, i} +  
\bra{\alpha +, i}\hat{t}_{ij}\ket{\beta -, j} \bra{\beta -, j}\hat{\vec{r}}_{ji}\ket{\alpha +, i} \right. \nonumber \\ 
&\left.  + \bra{\alpha +, j}\hat{\vec{r}}_{ji}\ket{\beta -, i} \bra{\beta -, i}\hat{t}_{ji}\ket{\alpha +, j} +  
\bra{\alpha +, j}\hat{t}_{ji}\ket{\beta -, i} \bra{\beta -, i}\hat{\vec{r}}_{ij}\ket{\alpha +, j} \right].
\end{align}
Similar to the hopping matrix, the position matrix can be decomposed into a linear combination of the Pauli matrices as:
\begin{align}
\label{eqn:R_pauli}
\hat{\vec{r}}_{ij}=  \hat{\vec{R}}^{0}_{ij} \otimes \sigma_0 + \sum_{\gamma'} i \hat{\vec{R}}^{\gamma'}_{ij}  \otimes \hat{\sigma}_{\gamma'}.
\end{align}
Then, it is straightforward to find the pair interaction formula Eq.~(\ref{eqn:P_s}) and the corresponding analytical formula Eq.~(\ref{eqn:Ps}).

\subsection{Single-$q$ spiral Ansatz}
In the simplest approximation, a spin-spiral state can be characterized by a single $q$-vector,
\begin{align}
\label{eq:Ans}
    \vec{e}_{i} = \hat{n}_1 \cos (\vec{q}\cdot \vec{R}_i) + \hat{n}_2 \sin (\vec{q}\cdot \vec{R}_i),
\end{align}
where $\hat{n}_1$ and $\hat{n}_2$ are orthogonal vectors within the plane of the spiral, $\vec{q}$ is the propagation vector and $\vec{R}_i$ is the vector pointing to the site $i$. 
This is an Ansatz of a single-$q$ spiral state.
The spiral plane can be characterized by its normal vector, 
\begin{align}
    \hat{n}^{\perp} = (-\sin \phi, \cos\phi, 0),
\end{align}
where $\phi$ indicates the rotation of the spin-spiral plane. 
In this case, the relative orientation between $\delta \vec{q}$ and $\hat{n}^{\perp}$ defines the spiral type. 
For instance, $\delta \vec{q} \parallel \hat{n}^{\perp}$ is a proper-screw spiral state. 

\subsection{Layer AFM chain continuous model}
We express magnetization and AFM order parameter at position $x$ as $\vec{M} (x) =\vec{S}_1(x) + \vec{S}_2(x)$ and $\vec{A}(x) = \vec{S}_1(x) - \vec{S}_2(x)$, respectively. 
In this formulas, $\vec{S}_1(x)$ and $\vec{S}_2(x)$ are spins at neighboring layer $L1$ and $L2$ and these satisfy $\vec{A}(x) \cdot \vec{M}(x) = 0$.
Then, energy density of the chain at position $x$ is expressed as 
\begin{align}
\label{eqn:Enden}
        \delta E(x) = & -\frac{J'}{4} (\nabla \vec{A}(x))^2  +  \frac{\vec{D}}{2} \cdot\left[  \vec{A}(x) \times \left( \nabla\vec{A}(x)\right) \right] - \vec{H} \cdot \vec{M}(x) + J.
\end{align}
In this equation, the first term is the effective isotropic ferromagnetic exchange, $J'<0$, the second one is the antisymmetric DM interaction, the third term corresponds to the Zeeman energy due to an external field $\vec{H}$, and the last term is the energy of a ferromagnetic state. 
The alternating inter-layer AFM exchange coupling in NISO can be considered as intra-layer FM exchanges. 
Considering inter-layer AFM exchange $J$, the effective intra-layer FM exchange would be $J' = -2J$.
By employing the Fourier expansion of the spin, $ \vec{A}(x) = \frac{1}{\sqrt{N}} \sum_{n=-N}^{N} \vec{A}_{n} \; e^{iQnx}$, we obtain a simplified energy expression that has to be minimized:
\begin{align}
\label{eqn:Enrec}
        E_{\mathrm{rec}} = & \sum_{n=-N}^{N} \left[ -\frac{J'}{4} Q^2 n^2 \vec{A}_{n}\cdot \vec{A}_{-n} + \frac{\vec{D}}{2} iQ n \cdot \left( \vec{A}_{-n} \times \vec{A}_{n} \right)\right] 
        - \left(\vec{H} \cdot \vec{A}_{0}\right)^2 \nonumber \\
        & + \lambda \left(\sum_{l'm'n'p'}  e^{iQ(l'+m'+n'+p')x} \vec{A}_{l'} \vec{A}_{m'} \vec{A}_{n'} \vec{A}_{p'} -2 \vec{A}_{n} \vec{A}_{-n}\right),
\end{align}
 where the last term is the energy penalty with coefficient $\lambda\gg 1$ enforcing a constraint on the spin length.
 Similarly, the change of the spin-isotropic part of the polarization per spin can be evaluated as:
\begin{align}
             P^{\mathrm{iso}}_z \propto -1 +  \sum_{n=-N}^{N} Q^2 n^2 \vec{A}_{n}\cdot \vec{A}_{-n}.
\end{align} 

\section{Data availability statement}
The data that used in this study are available
from the corresponding author upon reasonable request.

\section{Acknowledgements}
We thank M. Mostovoy and S. A. Nikolaev for fruitful discussions. R.O, was supported by JSPS KAKENHI Grant Numbers JP23KJ2165.

\section{Competing interests}
The authors declare no competing interests.

\section{Author Contributions}
R.O. performed first principles calculations, analytical formulas derivations, MCS simulations and 1D continuous model simulations. R.O. and I.S. performed derivation of the multiplet energy in the superexchange theory. R.O. and S.A. performed derivation of the 1D continuous model and the Yukawa-like potential. All authors participated in analysing the data and discussions. R.O. wrote the main manuscript text with contributions from all authors.

\section{References}
\bibliography{ref}

\begin{thebibliography}{43}%
\makeatletter
\providecommand \@ifxundefined [1]{%
 \@ifx{#1\undefined}
}%
\providecommand \@ifnum [1]{%
 \ifnum #1\expandafter \@firstoftwo
 \else \expandafter \@secondoftwo
 \fi
}%
\providecommand \@ifx [1]{%
 \ifx #1\expandafter \@firstoftwo
 \else \expandafter \@secondoftwo
 \fi
}%
\providecommand \natexlab [1]{#1}%
\providecommand \enquote  [1]{``#1''}%
\providecommand \bibnamefont  [1]{#1}%
\providecommand \bibfnamefont [1]{#1}%
\providecommand \citenamefont [1]{#1}%
\providecommand \href@noop [0]{\@secondoftwo}%
\providecommand \href [0]{\begingroup \@sanitize@url \@href}%
\providecommand \@href[1]{\@@startlink{#1}\@@href}%
\providecommand \@@href[1]{\endgroup#1\@@endlink}%
\providecommand \@sanitize@url [0]{\catcode `\\12\catcode `\$12\catcode
  `\&12\catcode `\#12\catcode `\^12\catcode `\_12\catcode `\%12\relax}%
\providecommand \@@startlink[1]{}%
\providecommand \@@endlink[0]{}%
\providecommand \url  [0]{\begingroup\@sanitize@url \@url }%
\providecommand \@url [1]{\endgroup\@href {#1}{\urlprefix }}%
\providecommand \urlprefix  [0]{URL }%
\providecommand \Eprint [0]{\href }%
\providecommand \doibase [0]{https://doi.org/}%
\providecommand \selectlanguage [0]{\@gobble}%
\providecommand \bibinfo  [0]{\@secondoftwo}%
\providecommand \bibfield  [0]{\@secondoftwo}%
\providecommand \translation [1]{[#1]}%
\providecommand \BibitemOpen [0]{}%
\providecommand \bibitemStop [0]{}%
\providecommand \bibitemNoStop [0]{.\EOS\space}%
\providecommand \EOS [0]{\spacefactor3000\relax}%
\providecommand \BibitemShut  [1]{\csname bibitem#1\endcsname}%
\let\auto@bib@innerbib\@empty
\bibitem [{\citenamefont {Kimura}\ \emph {et~al.}(2003)\citenamefont {Kimura},
  \citenamefont {Goto}, \citenamefont {Shintani}, \citenamefont {Ishizaka},
  \citenamefont {Arima},\ and\ \citenamefont {Tokura}}]{Kimura2003}%
  \BibitemOpen
  \bibfield  {author} {\bibinfo {author} {\bibfnamefont {T.}~\bibnamefont
  {Kimura}}, \bibinfo {author} {\bibfnamefont {T.}~\bibnamefont {Goto}},
  \bibinfo {author} {\bibfnamefont {H.}~\bibnamefont {Shintani}}, \bibinfo
  {author} {\bibfnamefont {K.}~\bibnamefont {Ishizaka}}, \bibinfo {author}
  {\bibfnamefont {T.}~\bibnamefont {Arima}},\ and\ \bibinfo {author}
  {\bibfnamefont {Y.}~\bibnamefont {Tokura}},\ }\href
  {https://doi.org/10.1038/nature02018} {\bibfield  {journal} {\bibinfo
  {journal} {Nature}\ }\textbf {\bibinfo {volume} {426}},\ \bibinfo {pages}
  {55} (\bibinfo {year} {2003})}\BibitemShut {NoStop}%
\bibitem [{\citenamefont {Tokura}\ and\ \citenamefont
  {Seki}(2010)}]{Tokura_ad}%
  \BibitemOpen
  \bibfield  {author} {\bibinfo {author} {\bibfnamefont {Y.}~\bibnamefont
  {Tokura}}\ and\ \bibinfo {author} {\bibfnamefont {S.}~\bibnamefont {Seki}},\
  }\href {https://doi.org/https://doi.org/10.1002/adma.200901961} {\bibfield
  {journal} {\bibinfo  {journal} {Advanced Materials}\ }\textbf {\bibinfo
  {volume} {22}},\ \bibinfo {pages} {1554} (\bibinfo {year}
  {2010})}\BibitemShut {NoStop}%
\bibitem [{\citenamefont {Tokura}\ \emph {et~al.}(2014)\citenamefont {Tokura},
  \citenamefont {Seki},\ and\ \citenamefont {Nagaosa}}]{Tokura2014}%
  \BibitemOpen
  \bibfield  {author} {\bibinfo {author} {\bibfnamefont {Y.}~\bibnamefont
  {Tokura}}, \bibinfo {author} {\bibfnamefont {S.}~\bibnamefont {Seki}},\ and\
  \bibinfo {author} {\bibfnamefont {N.}~\bibnamefont {Nagaosa}},\ }\href
  {https://doi.org/10.1088/0034-4885/77/7/076501} {\bibfield  {journal}
  {\bibinfo  {journal} {Reports on Progress in Physics}\ }\textbf {\bibinfo
  {volume} {77}},\ \bibinfo {pages} {076501} (\bibinfo {year}
  {2014})}\BibitemShut {NoStop}%
\bibitem [{\citenamefont {Khomskii}(2009)}]{Khomskii2009classifying}%
  \BibitemOpen
  \bibfield  {author} {\bibinfo {author} {\bibfnamefont {D.}~\bibnamefont
  {Khomskii}},\ }\href@noop {} {\bibfield  {journal} {\bibinfo  {journal}
  {Physics}\ }\textbf {\bibinfo {volume} {2}},\ \bibinfo {pages} {20} (\bibinfo
  {year} {2009})}\BibitemShut {NoStop}%
\bibitem [{\citenamefont {Cheong}\ and\ \citenamefont
  {Mostovoy}(2007)}]{Cheong2007}%
  \BibitemOpen
  \bibfield  {author} {\bibinfo {author} {\bibfnamefont {S.-W.}\ \bibnamefont
  {Cheong}}\ and\ \bibinfo {author} {\bibfnamefont {M.}~\bibnamefont
  {Mostovoy}},\ }\href {https://doi.org/10.1038/nmat1804} {\bibfield  {journal}
  {\bibinfo  {journal} {Nature Materials}\ }\textbf {\bibinfo {volume} {6}},\
  \bibinfo {pages} {13} (\bibinfo {year} {2007})}\BibitemShut {NoStop}%
\bibitem [{\citenamefont {Fiebig}\ \emph {et~al.}(2016)\citenamefont {Fiebig},
  \citenamefont {Lottermoser}, \citenamefont {Meier},\ and\ \citenamefont
  {Trassin}}]{Fiebig_2016}%
  \BibitemOpen
  \bibfield  {author} {\bibinfo {author} {\bibfnamefont {M.}~\bibnamefont
  {Fiebig}}, \bibinfo {author} {\bibfnamefont {T.}~\bibnamefont {Lottermoser}},
  \bibinfo {author} {\bibfnamefont {D.}~\bibnamefont {Meier}},\ and\ \bibinfo
  {author} {\bibfnamefont {M.}~\bibnamefont {Trassin}},\ }\href
  {https://doi.org/10.1038/natrevmats.2016.46} {\bibfield  {journal} {\bibinfo
  {journal} {Nature Reviews Materials}\ }\textbf {\bibinfo {volume} {1}},\
  \bibinfo {pages} {16046} (\bibinfo {year} {2016})}\BibitemShut {NoStop}%
\bibitem [{\citenamefont {Seki}\ \emph {et~al.}(2012)\citenamefont {Seki},
  \citenamefont {Ishiwata},\ and\ \citenamefont {Tokura}}]{Cu2OSeO3}%
  \BibitemOpen
  \bibfield  {author} {\bibinfo {author} {\bibfnamefont {S.}~\bibnamefont
  {Seki}}, \bibinfo {author} {\bibfnamefont {S.}~\bibnamefont {Ishiwata}},\
  and\ \bibinfo {author} {\bibfnamefont {Y.}~\bibnamefont {Tokura}},\ }\href
  {https://doi.org/10.1103/PhysRevB.86.060403} {\bibfield  {journal} {\bibinfo
  {journal} {Phys. Rev. B}\ }\textbf {\bibinfo {volume} {86}},\ \bibinfo
  {pages} {060403} (\bibinfo {year} {2012})}\BibitemShut {NoStop}%
\bibitem [{\citenamefont {Ruff}\ \emph {et~al.}(2015)\citenamefont {Ruff},
  \citenamefont {Widmann}, \citenamefont {Lunkenheimer}, \citenamefont
  {Tsurkan}, \citenamefont {Bordács}, \citenamefont {Kézsmárki},\ and\
  \citenamefont {Loidl}}]{GVS}%
  \BibitemOpen
  \bibfield  {author} {\bibinfo {author} {\bibfnamefont {E.}~\bibnamefont
  {Ruff}}, \bibinfo {author} {\bibfnamefont {S.}~\bibnamefont {Widmann}},
  \bibinfo {author} {\bibfnamefont {P.}~\bibnamefont {Lunkenheimer}}, \bibinfo
  {author} {\bibfnamefont {V.}~\bibnamefont {Tsurkan}}, \bibinfo {author}
  {\bibfnamefont {S.}~\bibnamefont {Bordács}}, \bibinfo {author}
  {\bibfnamefont {I.}~\bibnamefont {Kézsmárki}},\ and\ \bibinfo {author}
  {\bibfnamefont {A.}~\bibnamefont {Loidl}},\ }\href
  {https://doi.org/10.1126/sciadv.1500916} {\bibfield  {journal} {\bibinfo
  {journal} {Science Advances}\ }\textbf {\bibinfo {volume} {1}},\ \bibinfo
  {pages} {e1500916} (\bibinfo {year} {2015})}\BibitemShut {NoStop}%
\bibitem [{\citenamefont {Ivanov}\ \emph {et~al.}(2013)\citenamefont {Ivanov},
  \citenamefont {Mathieu}, \citenamefont {Nordblad}, \citenamefont {Tellgren},
  \citenamefont {Ritter}, \citenamefont {Politova}, \citenamefont {Kaleva},
  \citenamefont {Mosunov}, \citenamefont {Stefanovich},\ and\ \citenamefont
  {Weil}}]{NISO_chem}%
  \BibitemOpen
  \bibfield  {author} {\bibinfo {author} {\bibfnamefont {S.~A.}\ \bibnamefont
  {Ivanov}}, \bibinfo {author} {\bibfnamefont {R.}~\bibnamefont {Mathieu}},
  \bibinfo {author} {\bibfnamefont {P.}~\bibnamefont {Nordblad}}, \bibinfo
  {author} {\bibfnamefont {R.}~\bibnamefont {Tellgren}}, \bibinfo {author}
  {\bibfnamefont {C.}~\bibnamefont {Ritter}}, \bibinfo {author} {\bibfnamefont
  {E.}~\bibnamefont {Politova}}, \bibinfo {author} {\bibfnamefont
  {G.}~\bibnamefont {Kaleva}}, \bibinfo {author} {\bibfnamefont
  {A.}~\bibnamefont {Mosunov}}, \bibinfo {author} {\bibfnamefont
  {S.}~\bibnamefont {Stefanovich}},\ and\ \bibinfo {author} {\bibfnamefont
  {M.}~\bibnamefont {Weil}},\ }\href {https://doi.org/10.1021/cm304095s}
  {\bibfield  {journal} {\bibinfo  {journal} {Chemistry of Materials}\ }\textbf
  {\bibinfo {volume} {25}},\ \bibinfo {pages} {935} (\bibinfo {year}
  {2013})}\BibitemShut {NoStop}%
\bibitem [{\citenamefont {Weil}\ \emph {et~al.}(2014)\citenamefont {Weil},
  \citenamefont {Mathieu}, \citenamefont {Nordblad},\ and\ \citenamefont
  {Ivanov}}]{NISO_crys}%
  \BibitemOpen
  \bibfield  {author} {\bibinfo {author} {\bibfnamefont {M.}~\bibnamefont
  {Weil}}, \bibinfo {author} {\bibfnamefont {R.}~\bibnamefont {Mathieu}},
  \bibinfo {author} {\bibfnamefont {P.}~\bibnamefont {Nordblad}},\ and\
  \bibinfo {author} {\bibfnamefont {S.}~\bibnamefont {Ivanov}},\ }\href@noop {}
  {\bibfield  {journal} {\bibinfo  {journal} {Crystal Research and Technology}\
  }\textbf {\bibinfo {volume} {49}},\ \bibinfo {pages} {142} (\bibinfo {year}
  {2014})}\BibitemShut {NoStop}%
\bibitem [{\citenamefont {Kim}\ \emph {et~al.}(2015)\citenamefont {Kim},
  \citenamefont {Artyukhin}, \citenamefont {Mun}, \citenamefont {Jaime},
  \citenamefont {Harrison}, \citenamefont {Hansen}, \citenamefont {Yang},
  \citenamefont {Oh}, \citenamefont {Vanderbilt}, \citenamefont {Zapf},\ and\
  \citenamefont {Cheong}}]{SergeyNTO1}%
  \BibitemOpen
  \bibfield  {author} {\bibinfo {author} {\bibfnamefont {J.~W.}\ \bibnamefont
  {Kim}}, \bibinfo {author} {\bibfnamefont {S.}~\bibnamefont {Artyukhin}},
  \bibinfo {author} {\bibfnamefont {E.~D.}\ \bibnamefont {Mun}}, \bibinfo
  {author} {\bibfnamefont {M.}~\bibnamefont {Jaime}}, \bibinfo {author}
  {\bibfnamefont {N.}~\bibnamefont {Harrison}}, \bibinfo {author}
  {\bibfnamefont {A.}~\bibnamefont {Hansen}}, \bibinfo {author} {\bibfnamefont
  {J.~J.}\ \bibnamefont {Yang}}, \bibinfo {author} {\bibfnamefont {Y.~S.}\
  \bibnamefont {Oh}}, \bibinfo {author} {\bibfnamefont {D.}~\bibnamefont
  {Vanderbilt}}, \bibinfo {author} {\bibfnamefont {V.~S.}\ \bibnamefont
  {Zapf}},\ and\ \bibinfo {author} {\bibfnamefont {S.-W.}\ \bibnamefont
  {Cheong}},\ }\href {https://doi.org/10.1103/PhysRevLett.115.137201}
  {\bibfield  {journal} {\bibinfo  {journal} {Phys. Rev. Lett.}\ }\textbf
  {\bibinfo {volume} {115}},\ \bibinfo {pages} {137201} (\bibinfo {year}
  {2015})}\BibitemShut {NoStop}%
\bibitem [{\citenamefont {Yokosuk}\ \emph {et~al.}(2016)\citenamefont
  {Yokosuk}, \citenamefont {al~Wahish}, \citenamefont {Artyukhin},
  \citenamefont {O'Neal}, \citenamefont {Mazumdar}, \citenamefont {Chen},
  \citenamefont {Yang}, \citenamefont {Oh}, \citenamefont {McGill},
  \citenamefont {Haule}, \citenamefont {Cheong}, \citenamefont {Vanderbilt},\
  and\ \citenamefont {Musfeldt}}]{SergeyNTO2}%
  \BibitemOpen
  \bibfield  {author} {\bibinfo {author} {\bibfnamefont {M.~O.}\ \bibnamefont
  {Yokosuk}}, \bibinfo {author} {\bibfnamefont {A.}~\bibnamefont {al~Wahish}},
  \bibinfo {author} {\bibfnamefont {S.}~\bibnamefont {Artyukhin}}, \bibinfo
  {author} {\bibfnamefont {K.~R.}\ \bibnamefont {O'Neal}}, \bibinfo {author}
  {\bibfnamefont {D.}~\bibnamefont {Mazumdar}}, \bibinfo {author}
  {\bibfnamefont {P.}~\bibnamefont {Chen}}, \bibinfo {author} {\bibfnamefont
  {J.}~\bibnamefont {Yang}}, \bibinfo {author} {\bibfnamefont {Y.~S.}\
  \bibnamefont {Oh}}, \bibinfo {author} {\bibfnamefont {S.~A.}\ \bibnamefont
  {McGill}}, \bibinfo {author} {\bibfnamefont {K.}~\bibnamefont {Haule}},
  \bibinfo {author} {\bibfnamefont {S.-W.}\ \bibnamefont {Cheong}}, \bibinfo
  {author} {\bibfnamefont {D.}~\bibnamefont {Vanderbilt}},\ and\ \bibinfo
  {author} {\bibfnamefont {J.~L.}\ \bibnamefont {Musfeldt}},\ }\href
  {https://doi.org/10.1103/PhysRevLett.117.147402} {\bibfield  {journal}
  {\bibinfo  {journal} {Phys. Rev. Lett.}\ }\textbf {\bibinfo {volume} {117}},\
  \bibinfo {pages} {147402} (\bibinfo {year} {2016})}\BibitemShut {NoStop}%
\bibitem [{\citenamefont {Cai}\ \emph {et~al.}(2017)\citenamefont {Cai},
  \citenamefont {Greenblatt},\ and\ \citenamefont {Li}}]{Corundum}%
  \BibitemOpen
  \bibfield  {author} {\bibinfo {author} {\bibfnamefont {G.-H.}\ \bibnamefont
  {Cai}}, \bibinfo {author} {\bibfnamefont {M.}~\bibnamefont {Greenblatt}},\
  and\ \bibinfo {author} {\bibfnamefont {M.-R.}\ \bibnamefont {Li}},\ }\href
  {https://doi.org/10.1021/acs.chemmater.7b01567} {\bibfield  {journal}
  {\bibinfo  {journal} {Chemistry of Materials}\ }\textbf {\bibinfo {volume}
  {29}},\ \bibinfo {pages} {5447} (\bibinfo {year} {2017})}\BibitemShut
  {NoStop}%
\bibitem [{\citenamefont {Frank}\ \emph {et~al.}(2022)\citenamefont {Frank},
  \citenamefont {McCabe}, \citenamefont {Orlandi}, \citenamefont {Manuel},
  \citenamefont {Tan}, \citenamefont {Deng}, \citenamefont {Jin}, \citenamefont
  {Croft}, \citenamefont {Emge}, \citenamefont {Yu}, \citenamefont {Wang},
  \citenamefont {Gopalan}, \citenamefont {Lapidus}, \citenamefont {Wu},
  \citenamefont {Li}, \citenamefont {Gross}, \citenamefont {Burger},
  \citenamefont {Mielewczyk-Gry{\'{n}}}, \citenamefont {Klimczuk},
  \citenamefont {Xie}, \citenamefont {Walker},\ and\ \citenamefont
  {Greenblatt}}]{Frank2022}%
  \BibitemOpen
  \bibfield  {author} {\bibinfo {author} {\bibfnamefont {C.~E.}\ \bibnamefont
  {Frank}}, \bibinfo {author} {\bibfnamefont {E.~E.}\ \bibnamefont {McCabe}},
  \bibinfo {author} {\bibfnamefont {F.}~\bibnamefont {Orlandi}}, \bibinfo
  {author} {\bibfnamefont {P.}~\bibnamefont {Manuel}}, \bibinfo {author}
  {\bibfnamefont {X.}~\bibnamefont {Tan}}, \bibinfo {author} {\bibfnamefont
  {Z.}~\bibnamefont {Deng}}, \bibinfo {author} {\bibfnamefont {C.}~\bibnamefont
  {Jin}}, \bibinfo {author} {\bibfnamefont {M.}~\bibnamefont {Croft}}, \bibinfo
  {author} {\bibfnamefont {T.}~\bibnamefont {Emge}}, \bibinfo {author}
  {\bibfnamefont {S.}~\bibnamefont {Yu}}, \bibinfo {author} {\bibfnamefont
  {H.}~\bibnamefont {Wang}}, \bibinfo {author} {\bibfnamefont {V.}~\bibnamefont
  {Gopalan}}, \bibinfo {author} {\bibfnamefont {S.}~\bibnamefont {Lapidus}},
  \bibinfo {author} {\bibfnamefont {M.}~\bibnamefont {Wu}}, \bibinfo {author}
  {\bibfnamefont {M.-R.}\ \bibnamefont {Li}}, \bibinfo {author} {\bibfnamefont
  {J.}~\bibnamefont {Gross}}, \bibinfo {author} {\bibfnamefont
  {P.}~\bibnamefont {Burger}}, \bibinfo {author} {\bibfnamefont
  {A.}~\bibnamefont {Mielewczyk-Gry{\'{n}}}}, \bibinfo {author} {\bibfnamefont
  {T.}~\bibnamefont {Klimczuk}}, \bibinfo {author} {\bibfnamefont
  {W.}~\bibnamefont {Xie}}, \bibinfo {author} {\bibfnamefont {D.}~\bibnamefont
  {Walker}},\ and\ \bibinfo {author} {\bibfnamefont {M.}~\bibnamefont
  {Greenblatt}},\ }\href {https://doi.org/10.1021/acs.chemmater.2c00312}
  {\bibfield  {journal} {\bibinfo  {journal} {Chemistry of Materials}\ }\textbf
  {\bibinfo {volume} {34}},\ \bibinfo {pages} {5020} (\bibinfo {year}
  {2022})}\BibitemShut {NoStop}%
\bibitem [{\citenamefont {Ye}\ and\ \citenamefont
  {Vanderbilt}(2016)}]{Corundum_Vanderbilt}%
  \BibitemOpen
  \bibfield  {author} {\bibinfo {author} {\bibfnamefont {M.}~\bibnamefont
  {Ye}}\ and\ \bibinfo {author} {\bibfnamefont {D.}~\bibnamefont
  {Vanderbilt}},\ }\href {https://doi.org/10.1103/PhysRevB.93.134303}
  {\bibfield  {journal} {\bibinfo  {journal} {Phys. Rev. B}\ }\textbf {\bibinfo
  {volume} {93}},\ \bibinfo {pages} {134303} (\bibinfo {year}
  {2016})}\BibitemShut {NoStop}%
\bibitem [{\citenamefont {Araki}\ \emph {et~al.}(2020)\citenamefont {Araki},
  \citenamefont {Sato}, \citenamefont {Fujima}, \citenamefont {Abe},
  \citenamefont {Tokunaga}, \citenamefont {Kimura}, \citenamefont {Morikawa},
  \citenamefont {Ukleev}, \citenamefont {Yamasaki}, \citenamefont {Tabata},
  \citenamefont {Nakao}, \citenamefont {Murakami}, \citenamefont {Sagayama},
  \citenamefont {Ohishi}, \citenamefont {Tokunaga},\ and\ \citenamefont
  {Arima}}]{PRB_NISO}%
  \BibitemOpen
  \bibfield  {author} {\bibinfo {author} {\bibfnamefont {Y.}~\bibnamefont
  {Araki}}, \bibinfo {author} {\bibfnamefont {T.}~\bibnamefont {Sato}},
  \bibinfo {author} {\bibfnamefont {Y.}~\bibnamefont {Fujima}}, \bibinfo
  {author} {\bibfnamefont {N.}~\bibnamefont {Abe}}, \bibinfo {author}
  {\bibfnamefont {M.}~\bibnamefont {Tokunaga}}, \bibinfo {author}
  {\bibfnamefont {S.}~\bibnamefont {Kimura}}, \bibinfo {author} {\bibfnamefont
  {D.}~\bibnamefont {Morikawa}}, \bibinfo {author} {\bibfnamefont
  {V.}~\bibnamefont {Ukleev}}, \bibinfo {author} {\bibfnamefont
  {Y.}~\bibnamefont {Yamasaki}}, \bibinfo {author} {\bibfnamefont
  {C.}~\bibnamefont {Tabata}}, \bibinfo {author} {\bibfnamefont
  {H.}~\bibnamefont {Nakao}}, \bibinfo {author} {\bibfnamefont
  {Y.}~\bibnamefont {Murakami}}, \bibinfo {author} {\bibfnamefont
  {H.}~\bibnamefont {Sagayama}}, \bibinfo {author} {\bibfnamefont
  {K.}~\bibnamefont {Ohishi}}, \bibinfo {author} {\bibfnamefont
  {Y.}~\bibnamefont {Tokunaga}},\ and\ \bibinfo {author} {\bibfnamefont
  {T.}~\bibnamefont {Arima}},\ }\href
  {https://doi.org/10.1103/PhysRevB.102.054409} {\bibfield  {journal} {\bibinfo
   {journal} {Phys. Rev. B}\ }\textbf {\bibinfo {volume} {102}},\ \bibinfo
  {pages} {054409} (\bibinfo {year} {2020})}\BibitemShut {NoStop}%
\bibitem [{\citenamefont {Liu}\ \emph {et~al.}(2023)\citenamefont {Liu},
  \citenamefont {Araki}, \citenamefont {Arima}, \citenamefont {Itoh},
  \citenamefont {Asai},\ and\ \citenamefont {Masuda}}]{PRB_NISO_2}%
  \BibitemOpen
  \bibfield  {author} {\bibinfo {author} {\bibfnamefont {Z.}~\bibnamefont
  {Liu}}, \bibinfo {author} {\bibfnamefont {Y.}~\bibnamefont {Araki}}, \bibinfo
  {author} {\bibfnamefont {T.-h.}\ \bibnamefont {Arima}}, \bibinfo {author}
  {\bibfnamefont {S.}~\bibnamefont {Itoh}}, \bibinfo {author} {\bibfnamefont
  {S.}~\bibnamefont {Asai}},\ and\ \bibinfo {author} {\bibfnamefont
  {T.}~\bibnamefont {Masuda}},\ }\href
  {https://doi.org/10.1103/PhysRevB.107.064428} {\bibfield  {journal} {\bibinfo
   {journal} {Phys. Rev. B}\ }\textbf {\bibinfo {volume} {107}},\ \bibinfo
  {pages} {064428} (\bibinfo {year} {2023})}\BibitemShut {NoStop}%
\bibitem [{\citenamefont {Ihara}\ \emph {et~al.}(2023)\citenamefont {Ihara},
  \citenamefont {Hiyoshi}, \citenamefont {Shimohashi}, \citenamefont {Kumar},
  \citenamefont {Sasaki}, \citenamefont {Hirata}, \citenamefont {Araki},
  \citenamefont {Tokunaga},\ and\ \citenamefont {Arima}}]{PRB_NISO_3}%
  \BibitemOpen
  \bibfield  {author} {\bibinfo {author} {\bibfnamefont {Y.}~\bibnamefont
  {Ihara}}, \bibinfo {author} {\bibfnamefont {R.}~\bibnamefont {Hiyoshi}},
  \bibinfo {author} {\bibfnamefont {M.}~\bibnamefont {Shimohashi}}, \bibinfo
  {author} {\bibfnamefont {R.}~\bibnamefont {Kumar}}, \bibinfo {author}
  {\bibfnamefont {T.}~\bibnamefont {Sasaki}}, \bibinfo {author} {\bibfnamefont
  {M.}~\bibnamefont {Hirata}}, \bibinfo {author} {\bibfnamefont
  {Y.}~\bibnamefont {Araki}}, \bibinfo {author} {\bibfnamefont
  {Y.}~\bibnamefont {Tokunaga}},\ and\ \bibinfo {author} {\bibfnamefont
  {T.}~\bibnamefont {Arima}},\ }\href
  {https://doi.org/10.1103/PhysRevB.108.024417} {\bibfield  {journal} {\bibinfo
   {journal} {Phys. Rev. B}\ }\textbf {\bibinfo {volume} {108}},\ \bibinfo
  {pages} {024417} (\bibinfo {year} {2023})}\BibitemShut {NoStop}%
\bibitem [{\citenamefont {Choi}\ \emph {et~al.}(2008)\citenamefont {Choi},
  \citenamefont {Yi}, \citenamefont {Lee}, \citenamefont {Huang}, \citenamefont
  {Kiryukhin},\ and\ \citenamefont {Cheong}}]{strict}%
  \BibitemOpen
  \bibfield  {author} {\bibinfo {author} {\bibfnamefont {Y.~J.}\ \bibnamefont
  {Choi}}, \bibinfo {author} {\bibfnamefont {H.~T.}\ \bibnamefont {Yi}},
  \bibinfo {author} {\bibfnamefont {S.}~\bibnamefont {Lee}}, \bibinfo {author}
  {\bibfnamefont {Q.}~\bibnamefont {Huang}}, \bibinfo {author} {\bibfnamefont
  {V.}~\bibnamefont {Kiryukhin}},\ and\ \bibinfo {author} {\bibfnamefont
  {S.-W.}\ \bibnamefont {Cheong}},\ }\href
  {https://doi.org/10.1103/PhysRevLett.100.047601} {\bibfield  {journal}
  {\bibinfo  {journal} {Phys. Rev. Lett.}\ }\textbf {\bibinfo {volume} {100}},\
  \bibinfo {pages} {047601} (\bibinfo {year} {2008})}\BibitemShut {NoStop}%
\bibitem [{\citenamefont {Katsura}\ \emph {et~al.}(2005)\citenamefont
  {Katsura}, \citenamefont {Nagaosa},\ and\ \citenamefont {Balatsky}}]{KNB}%
  \BibitemOpen
  \bibfield  {author} {\bibinfo {author} {\bibfnamefont {H.}~\bibnamefont
  {Katsura}}, \bibinfo {author} {\bibfnamefont {N.}~\bibnamefont {Nagaosa}},\
  and\ \bibinfo {author} {\bibfnamefont {A.~V.}\ \bibnamefont {Balatsky}},\
  }\href {https://doi.org/10.1103/PhysRevLett.95.057205} {\bibfield  {journal}
  {\bibinfo  {journal} {Phys. Rev. Lett.}\ }\textbf {\bibinfo {volume} {95}},\
  \bibinfo {pages} {057205} (\bibinfo {year} {2005})}\BibitemShut {NoStop}%
\bibitem [{\citenamefont {Arima}(2007)}]{Arima}%
  \BibitemOpen
  \bibfield  {author} {\bibinfo {author} {\bibfnamefont {T.}~\bibnamefont
  {Arima}},\ }\href {https://doi.org/10.1143/JPSJ.76.073702} {\bibfield
  {journal} {\bibinfo  {journal} {Journal of the Physical Society of Japan}\
  }\textbf {\bibinfo {volume} {76}},\ \bibinfo {pages} {073702} (\bibinfo
  {year} {2007})}\BibitemShut {NoStop}%
\bibitem [{\citenamefont {King-Smith}\ and\ \citenamefont
  {Vanderbilt}(1993)}]{King}%
  \BibitemOpen
  \bibfield  {author} {\bibinfo {author} {\bibfnamefont {R.~D.}\ \bibnamefont
  {King-Smith}}\ and\ \bibinfo {author} {\bibfnamefont {D.}~\bibnamefont
  {Vanderbilt}},\ }\href {https://doi.org/10.1103/PhysRevB.47.1651} {\bibfield
  {journal} {\bibinfo  {journal} {Phys. Rev. B}\ }\textbf {\bibinfo {volume}
  {47}},\ \bibinfo {pages} {1651} (\bibinfo {year} {1993})}\BibitemShut
  {NoStop}%
\bibitem [{\citenamefont {Vanderbilt}\ and\ \citenamefont
  {King-Smith}(1993)}]{Vdb}%
  \BibitemOpen
  \bibfield  {author} {\bibinfo {author} {\bibfnamefont {D.}~\bibnamefont
  {Vanderbilt}}\ and\ \bibinfo {author} {\bibfnamefont {R.~D.}\ \bibnamefont
  {King-Smith}},\ }\href {https://doi.org/10.1103/PhysRevB.48.4442} {\bibfield
  {journal} {\bibinfo  {journal} {Phys. Rev. B}\ }\textbf {\bibinfo {volume}
  {48}},\ \bibinfo {pages} {4442} (\bibinfo {year} {1993})}\BibitemShut
  {NoStop}%
\bibitem [{\citenamefont {Resta}(2010)}]{Resta}%
  \BibitemOpen
  \bibfield  {author} {\bibinfo {author} {\bibfnamefont {R.}~\bibnamefont
  {Resta}},\ }\href {https://doi.org/10.1088/0953-8984/22/12/123201} {\bibfield
   {journal} {\bibinfo  {journal} {Journal of Physics: Condensed Matter}\
  }\textbf {\bibinfo {volume} {22}},\ \bibinfo {pages} {123201} (\bibinfo
  {year} {2010})}\BibitemShut {NoStop}%
\bibitem [{\citenamefont {Perdew}\ \emph {et~al.}(1996)\citenamefont {Perdew},
  \citenamefont {Burke},\ and\ \citenamefont {Ernzerhof}}]{PBE}%
  \BibitemOpen
  \bibfield  {author} {\bibinfo {author} {\bibfnamefont {J.~P.}\ \bibnamefont
  {Perdew}}, \bibinfo {author} {\bibfnamefont {K.}~\bibnamefont {Burke}},\ and\
  \bibinfo {author} {\bibfnamefont {M.}~\bibnamefont {Ernzerhof}},\ }\href
  {https://doi.org/10.1103/PhysRevLett.77.3865} {\bibfield  {journal} {\bibinfo
   {journal} {Phys. Rev. Lett.}\ }\textbf {\bibinfo {volume} {77}},\ \bibinfo
  {pages} {3865} (\bibinfo {year} {1996})}\BibitemShut {NoStop}%
\bibitem [{\citenamefont {Perdew}\ \emph {et~al.}(1997)\citenamefont {Perdew},
  \citenamefont {Burke},\ and\ \citenamefont {Ernzerhof}}]{PBE2}%
  \BibitemOpen
  \bibfield  {author} {\bibinfo {author} {\bibfnamefont {J.~P.}\ \bibnamefont
  {Perdew}}, \bibinfo {author} {\bibfnamefont {K.}~\bibnamefont {Burke}},\ and\
  \bibinfo {author} {\bibfnamefont {M.}~\bibnamefont {Ernzerhof}},\ }\href
  {https://doi.org/10.1103/PhysRevLett.78.1396} {\bibfield  {journal} {\bibinfo
   {journal} {Phys. Rev. Lett.}\ }\textbf {\bibinfo {volume} {78}},\ \bibinfo
  {pages} {1396} (\bibinfo {year} {1997})}\BibitemShut {NoStop}%
\bibitem [{\citenamefont {Kanamori}(1963)}]{Kanamori}%
  \BibitemOpen
  \bibfield  {author} {\bibinfo {author} {\bibfnamefont {J.}~\bibnamefont
  {Kanamori}},\ }\href {https://doi.org/10.1143/PTP.30.275} {\bibfield
  {journal} {\bibinfo  {journal} {Progress of Theoretical Physics}\ }\textbf
  {\bibinfo {volume} {30}},\ \bibinfo {pages} {275} (\bibinfo {year}
  {1963})}\BibitemShut {NoStop}%
\bibitem [{\citenamefont {Moriya}(1960)}]{Moriya}%
  \BibitemOpen
  \bibfield  {author} {\bibinfo {author} {\bibfnamefont {T.}~\bibnamefont
  {Moriya}},\ }\href {https://doi.org/10.1103/PhysRev.120.91} {\bibfield
  {journal} {\bibinfo  {journal} {Phys. Rev.}\ }\textbf {\bibinfo {volume}
  {120}},\ \bibinfo {pages} {91} (\bibinfo {year} {1960})}\BibitemShut
  {NoStop}%
\bibitem [{\citenamefont {Nikolaev}\ and\ \citenamefont
  {Solovyev}(2019)}]{PRB_GVS}%
  \BibitemOpen
  \bibfield  {author} {\bibinfo {author} {\bibfnamefont {S.~A.}\ \bibnamefont
  {Nikolaev}}\ and\ \bibinfo {author} {\bibfnamefont {I.~V.}\ \bibnamefont
  {Solovyev}},\ }\href {https://doi.org/10.1103/PhysRevB.99.100401} {\bibfield
  {journal} {\bibinfo  {journal} {Phys. Rev. B}\ }\textbf {\bibinfo {volume}
  {99}},\ \bibinfo {pages} {100401} (\bibinfo {year} {2019})}\BibitemShut
  {NoStop}%
\bibitem [{\citenamefont {Ono}\ \emph {et~al.}(2020)\citenamefont {Ono},
  \citenamefont {Nikolaev},\ and\ \citenamefont {Solovyev}}]{PRB_RIS}%
  \BibitemOpen
  \bibfield  {author} {\bibinfo {author} {\bibfnamefont {R.}~\bibnamefont
  {Ono}}, \bibinfo {author} {\bibfnamefont {S.}~\bibnamefont {Nikolaev}},\ and\
  \bibinfo {author} {\bibfnamefont {I.}~\bibnamefont {Solovyev}},\ }\href
  {https://doi.org/10.1103/PhysRevB.102.064422} {\bibfield  {journal} {\bibinfo
   {journal} {Phys. Rev. B}\ }\textbf {\bibinfo {volume} {102}},\ \bibinfo
  {pages} {064422} (\bibinfo {year} {2020})}\BibitemShut {NoStop}%
\bibitem [{\citenamefont {Solovyev}\ \emph {et~al.}(2021)\citenamefont
  {Solovyev}, \citenamefont {Ono},\ and\ \citenamefont {Nikolaev}}]{PRL_IRS}%
  \BibitemOpen
  \bibfield  {author} {\bibinfo {author} {\bibfnamefont {I.}~\bibnamefont
  {Solovyev}}, \bibinfo {author} {\bibfnamefont {R.}~\bibnamefont {Ono}},\ and\
  \bibinfo {author} {\bibfnamefont {S.}~\bibnamefont {Nikolaev}},\ }\href
  {https://doi.org/10.1103/PhysRevLett.127.187601} {\bibfield  {journal}
  {\bibinfo  {journal} {Phys. Rev. Lett.}\ }\textbf {\bibinfo {volume} {127}},\
  \bibinfo {pages} {187601} (\bibinfo {year} {2021})}\BibitemShut {NoStop}%
\bibitem [{\citenamefont {Solovyev}(2015)}]{Igor_BCGO}%
  \BibitemOpen
  \bibfield  {author} {\bibinfo {author} {\bibfnamefont {I.~V.}\ \bibnamefont
  {Solovyev}},\ }\href {https://doi.org/10.1103/PhysRevB.91.224423} {\bibfield
  {journal} {\bibinfo  {journal} {Phys. Rev. B}\ }\textbf {\bibinfo {volume}
  {91}},\ \bibinfo {pages} {224423} (\bibinfo {year} {2015})}\BibitemShut
  {NoStop}%
\bibitem [{\citenamefont {Goodenough}(1955)}]{GK}%
  \BibitemOpen
  \bibfield  {author} {\bibinfo {author} {\bibfnamefont {J.~B.}\ \bibnamefont
  {Goodenough}},\ }\href {https://doi.org/10.1103/PhysRev.100.564} {\bibfield
  {journal} {\bibinfo  {journal} {Phys. Rev.}\ }\textbf {\bibinfo {volume}
  {100}},\ \bibinfo {pages} {564} (\bibinfo {year} {1955})}\BibitemShut
  {NoStop}%
\bibitem [{\citenamefont {Liechtenstein}\ \emph {et~al.}(1987)\citenamefont
  {Liechtenstein}, \citenamefont {Katsnelson}, \citenamefont {Antropov},\ and\
  \citenamefont {Gubanov}}]{Liechtenstein1987}%
  \BibitemOpen
  \bibfield  {author} {\bibinfo {author} {\bibfnamefont {A.}~\bibnamefont
  {Liechtenstein}}, \bibinfo {author} {\bibfnamefont {M.}~\bibnamefont
  {Katsnelson}}, \bibinfo {author} {\bibfnamefont {V.}~\bibnamefont
  {Antropov}},\ and\ \bibinfo {author} {\bibfnamefont {V.}~\bibnamefont
  {Gubanov}},\ }\href
  {https://doi.org/https://doi.org/10.1016/0304-8853(87)90721-9} {\bibfield
  {journal} {\bibinfo  {journal} {Journal of Magnetism and Magnetic Materials}\
  }\textbf {\bibinfo {volume} {67}},\ \bibinfo {pages} {65} (\bibinfo {year}
  {1987})}\BibitemShut {NoStop}%
\bibitem [{\citenamefont {{De Gennes}}(1968)}]{DeGennes}%
  \BibitemOpen
  \bibfield  {author} {\bibinfo {author} {\bibfnamefont {P.}~\bibnamefont {{De
  Gennes}}},\ }\href
  {https://doi.org/https://doi.org/10.1016/0038-1098(68)90024-0} {\bibfield
  {journal} {\bibinfo  {journal} {Solid State Communications}\ }\textbf
  {\bibinfo {volume} {6}},\ \bibinfo {pages} {163} (\bibinfo {year}
  {1968})}\BibitemShut {NoStop}%
\bibitem [{\citenamefont {Artyukhin}\ \emph {et~al.}(2012)\citenamefont
  {Artyukhin}, \citenamefont {Mostovoy}, \citenamefont {Jensen}, \citenamefont
  {Le}, \citenamefont {Prokes}, \citenamefont {de~Paula}, \citenamefont
  {Bordallo}, \citenamefont {Maljuk}, \citenamefont {Landsgesell},
  \citenamefont {Ryll}, \citenamefont {Klemke}, \citenamefont {Paeckel},
  \citenamefont {Kiefer}, \citenamefont {Lefmann}, \citenamefont {Kuhn},\ and\
  \citenamefont {Argyriou}}]{Artyukhin2012}%
  \BibitemOpen
  \bibfield  {author} {\bibinfo {author} {\bibfnamefont {S.}~\bibnamefont
  {Artyukhin}}, \bibinfo {author} {\bibfnamefont {M.}~\bibnamefont {Mostovoy}},
  \bibinfo {author} {\bibfnamefont {N.~P.}\ \bibnamefont {Jensen}}, \bibinfo
  {author} {\bibfnamefont {D.}~\bibnamefont {Le}}, \bibinfo {author}
  {\bibfnamefont {K.}~\bibnamefont {Prokes}}, \bibinfo {author} {\bibfnamefont
  {V.~G.}\ \bibnamefont {de~Paula}}, \bibinfo {author} {\bibfnamefont {H.~N.}\
  \bibnamefont {Bordallo}}, \bibinfo {author} {\bibfnamefont {A.}~\bibnamefont
  {Maljuk}}, \bibinfo {author} {\bibfnamefont {S.}~\bibnamefont {Landsgesell}},
  \bibinfo {author} {\bibfnamefont {H.}~\bibnamefont {Ryll}}, \bibinfo {author}
  {\bibfnamefont {B.}~\bibnamefont {Klemke}}, \bibinfo {author} {\bibfnamefont
  {S.}~\bibnamefont {Paeckel}}, \bibinfo {author} {\bibfnamefont
  {K.}~\bibnamefont {Kiefer}}, \bibinfo {author} {\bibfnamefont
  {K.}~\bibnamefont {Lefmann}}, \bibinfo {author} {\bibfnamefont {L.~T.}\
  \bibnamefont {Kuhn}},\ and\ \bibinfo {author} {\bibfnamefont {D.~N.}\
  \bibnamefont {Argyriou}},\ }\href {https://doi.org/10.1038/nmat3358}
  {\bibfield  {journal} {\bibinfo  {journal} {Nature Materials}\ }\textbf
  {\bibinfo {volume} {11}},\ \bibinfo {pages} {694} (\bibinfo {year}
  {2012})}\BibitemShut {NoStop}%
\bibitem [{\citenamefont {Giannozzi}\ \emph {et~al.}(2009)\citenamefont
  {Giannozzi}, \citenamefont {Baroni}, \citenamefont {Bonini}, \citenamefont
  {Calandra}, \citenamefont {Car}, \citenamefont {Cavazzoni}, \citenamefont
  {Ceresoli}, \citenamefont {Chiarotti}, \citenamefont {Cococcioni},
  \citenamefont {Dabo}, \citenamefont {Corso}, \citenamefont {de~Gironcoli},
  \citenamefont {Fabris}, \citenamefont {Fratesi}, \citenamefont {Gebauer},
  \citenamefont {Gerstmann}, \citenamefont {Gougoussis}, \citenamefont
  {Kokalj}, \citenamefont {Lazzeri}, \citenamefont {Martin-Samos},
  \citenamefont {Marzari}, \citenamefont {Mauri}, \citenamefont {Mazzarello},
  \citenamefont {Paolini}, \citenamefont {Pasquarello}, \citenamefont
  {Paulatto}, \citenamefont {Sbraccia}, \citenamefont {Scandolo}, \citenamefont
  {Sclauzero}, \citenamefont {Seitsonen}, \citenamefont {Smogunov},
  \citenamefont {Umari},\ and\ \citenamefont {Wentzcovitch}}]{QE1}%
  \BibitemOpen
  \bibfield  {author} {\bibinfo {author} {\bibfnamefont {P.}~\bibnamefont
  {Giannozzi}}, \bibinfo {author} {\bibfnamefont {S.}~\bibnamefont {Baroni}},
  \bibinfo {author} {\bibfnamefont {N.}~\bibnamefont {Bonini}}, \bibinfo
  {author} {\bibfnamefont {M.}~\bibnamefont {Calandra}}, \bibinfo {author}
  {\bibfnamefont {R.}~\bibnamefont {Car}}, \bibinfo {author} {\bibfnamefont
  {C.}~\bibnamefont {Cavazzoni}}, \bibinfo {author} {\bibfnamefont
  {D.}~\bibnamefont {Ceresoli}}, \bibinfo {author} {\bibfnamefont {G.~L.}\
  \bibnamefont {Chiarotti}}, \bibinfo {author} {\bibfnamefont {M.}~\bibnamefont
  {Cococcioni}}, \bibinfo {author} {\bibfnamefont {I.}~\bibnamefont {Dabo}},
  \bibinfo {author} {\bibfnamefont {A.~D.}\ \bibnamefont {Corso}}, \bibinfo
  {author} {\bibfnamefont {S.}~\bibnamefont {de~Gironcoli}}, \bibinfo {author}
  {\bibfnamefont {S.}~\bibnamefont {Fabris}}, \bibinfo {author} {\bibfnamefont
  {G.}~\bibnamefont {Fratesi}}, \bibinfo {author} {\bibfnamefont
  {R.}~\bibnamefont {Gebauer}}, \bibinfo {author} {\bibfnamefont
  {U.}~\bibnamefont {Gerstmann}}, \bibinfo {author} {\bibfnamefont
  {C.}~\bibnamefont {Gougoussis}}, \bibinfo {author} {\bibfnamefont
  {A.}~\bibnamefont {Kokalj}}, \bibinfo {author} {\bibfnamefont
  {M.}~\bibnamefont {Lazzeri}}, \bibinfo {author} {\bibfnamefont
  {L.}~\bibnamefont {Martin-Samos}}, \bibinfo {author} {\bibfnamefont
  {N.}~\bibnamefont {Marzari}}, \bibinfo {author} {\bibfnamefont
  {F.}~\bibnamefont {Mauri}}, \bibinfo {author} {\bibfnamefont
  {R.}~\bibnamefont {Mazzarello}}, \bibinfo {author} {\bibfnamefont
  {S.}~\bibnamefont {Paolini}}, \bibinfo {author} {\bibfnamefont
  {A.}~\bibnamefont {Pasquarello}}, \bibinfo {author} {\bibfnamefont
  {L.}~\bibnamefont {Paulatto}}, \bibinfo {author} {\bibfnamefont
  {C.}~\bibnamefont {Sbraccia}}, \bibinfo {author} {\bibfnamefont
  {S.}~\bibnamefont {Scandolo}}, \bibinfo {author} {\bibfnamefont
  {G.}~\bibnamefont {Sclauzero}}, \bibinfo {author} {\bibfnamefont {A.~P.}\
  \bibnamefont {Seitsonen}}, \bibinfo {author} {\bibfnamefont {A.}~\bibnamefont
  {Smogunov}}, \bibinfo {author} {\bibfnamefont {P.}~\bibnamefont {Umari}},\
  and\ \bibinfo {author} {\bibfnamefont {R.~M.}\ \bibnamefont {Wentzcovitch}},\
  }\href {https://doi.org/10.1088/0953-8984/21/39/395502} {\bibfield  {journal}
  {\bibinfo  {journal} {Journal of Physics: Condensed Matter}\ }\textbf
  {\bibinfo {volume} {21}},\ \bibinfo {pages} {395502} (\bibinfo {year}
  {2009})}\BibitemShut {NoStop}%
\bibitem [{\citenamefont {Marzari}\ and\ \citenamefont
  {Vanderbilt}(1997)}]{Marzari}%
  \BibitemOpen
  \bibfield  {author} {\bibinfo {author} {\bibfnamefont {N.}~\bibnamefont
  {Marzari}}\ and\ \bibinfo {author} {\bibfnamefont {D.}~\bibnamefont
  {Vanderbilt}},\ }\href {https://doi.org/10.1103/PhysRevB.56.12847} {\bibfield
   {journal} {\bibinfo  {journal} {Phys. Rev. B}\ }\textbf {\bibinfo {volume}
  {56}},\ \bibinfo {pages} {12847} (\bibinfo {year} {1997})}\BibitemShut
  {NoStop}%
\bibitem [{\citenamefont {Pizzi}\ \emph {et~al.}(2020)\citenamefont {Pizzi},
  \citenamefont {Vitale}, \citenamefont {Arita}, \citenamefont {Blügel},
  \citenamefont {Freimuth}, \citenamefont {Géranton}, \citenamefont
  {Gibertini}, \citenamefont {Gresch}, \citenamefont {Johnson}, \citenamefont
  {Koretsune}, \citenamefont {Ibañez-Azpiroz}, \citenamefont {Lee},
  \citenamefont {Lihm}, \citenamefont {Marchand}, \citenamefont {Marrazzo},
  \citenamefont {Mokrousov}, \citenamefont {Mustafa}, \citenamefont {Nohara},
  \citenamefont {Nomura}, \citenamefont {Paulatto}, \citenamefont {Poncé},
  \citenamefont {Ponweiser}, \citenamefont {Qiao}, \citenamefont {Thöle},
  \citenamefont {Tsirkin}, \citenamefont {Wierzbowska}, \citenamefont
  {Marzari}, \citenamefont {Vanderbilt}, \citenamefont {Souza}, \citenamefont
  {Mostofi},\ and\ \citenamefont {Yates}}]{W90}%
  \BibitemOpen
  \bibfield  {author} {\bibinfo {author} {\bibfnamefont {G.}~\bibnamefont
  {Pizzi}}, \bibinfo {author} {\bibfnamefont {V.}~\bibnamefont {Vitale}},
  \bibinfo {author} {\bibfnamefont {R.}~\bibnamefont {Arita}}, \bibinfo
  {author} {\bibfnamefont {S.}~\bibnamefont {Blügel}}, \bibinfo {author}
  {\bibfnamefont {F.}~\bibnamefont {Freimuth}}, \bibinfo {author}
  {\bibfnamefont {G.}~\bibnamefont {Géranton}}, \bibinfo {author}
  {\bibfnamefont {M.}~\bibnamefont {Gibertini}}, \bibinfo {author}
  {\bibfnamefont {D.}~\bibnamefont {Gresch}}, \bibinfo {author} {\bibfnamefont
  {C.}~\bibnamefont {Johnson}}, \bibinfo {author} {\bibfnamefont
  {T.}~\bibnamefont {Koretsune}}, \bibinfo {author} {\bibfnamefont
  {J.}~\bibnamefont {Ibañez-Azpiroz}}, \bibinfo {author} {\bibfnamefont
  {H.}~\bibnamefont {Lee}}, \bibinfo {author} {\bibfnamefont {J.-M.}\
  \bibnamefont {Lihm}}, \bibinfo {author} {\bibfnamefont {D.}~\bibnamefont
  {Marchand}}, \bibinfo {author} {\bibfnamefont {A.}~\bibnamefont {Marrazzo}},
  \bibinfo {author} {\bibfnamefont {Y.}~\bibnamefont {Mokrousov}}, \bibinfo
  {author} {\bibfnamefont {J.~I.}\ \bibnamefont {Mustafa}}, \bibinfo {author}
  {\bibfnamefont {Y.}~\bibnamefont {Nohara}}, \bibinfo {author} {\bibfnamefont
  {Y.}~\bibnamefont {Nomura}}, \bibinfo {author} {\bibfnamefont
  {L.}~\bibnamefont {Paulatto}}, \bibinfo {author} {\bibfnamefont
  {S.}~\bibnamefont {Poncé}}, \bibinfo {author} {\bibfnamefont
  {T.}~\bibnamefont {Ponweiser}}, \bibinfo {author} {\bibfnamefont
  {J.}~\bibnamefont {Qiao}}, \bibinfo {author} {\bibfnamefont {F.}~\bibnamefont
  {Thöle}}, \bibinfo {author} {\bibfnamefont {S.~S.}\ \bibnamefont {Tsirkin}},
  \bibinfo {author} {\bibfnamefont {M.}~\bibnamefont {Wierzbowska}}, \bibinfo
  {author} {\bibfnamefont {N.}~\bibnamefont {Marzari}}, \bibinfo {author}
  {\bibfnamefont {D.}~\bibnamefont {Vanderbilt}}, \bibinfo {author}
  {\bibfnamefont {I.}~\bibnamefont {Souza}}, \bibinfo {author} {\bibfnamefont
  {A.~A.}\ \bibnamefont {Mostofi}},\ and\ \bibinfo {author} {\bibfnamefont
  {J.~R.}\ \bibnamefont {Yates}},\ }\href
  {https://doi.org/10.1088/1361-648X/ab51ff} {\bibfield  {journal} {\bibinfo
  {journal} {Journal of Physics: Condensed Matter}\ }\textbf {\bibinfo {volume}
  {32}},\ \bibinfo {pages} {165902} (\bibinfo {year} {2020})}\BibitemShut
  {NoStop}%
\bibitem [{\citenamefont {Mostofi}\ \emph {et~al.}(2014)\citenamefont
  {Mostofi}, \citenamefont {Yates}, \citenamefont {Pizzi}, \citenamefont {Lee},
  \citenamefont {Souza}, \citenamefont {Vanderbilt},\ and\ \citenamefont
  {Marzari}}]{W90_2}%
  \BibitemOpen
  \bibfield  {author} {\bibinfo {author} {\bibfnamefont {A.~A.}\ \bibnamefont
  {Mostofi}}, \bibinfo {author} {\bibfnamefont {J.~R.}\ \bibnamefont {Yates}},
  \bibinfo {author} {\bibfnamefont {G.}~\bibnamefont {Pizzi}}, \bibinfo
  {author} {\bibfnamefont {Y.-S.}\ \bibnamefont {Lee}}, \bibinfo {author}
  {\bibfnamefont {I.}~\bibnamefont {Souza}}, \bibinfo {author} {\bibfnamefont
  {D.}~\bibnamefont {Vanderbilt}},\ and\ \bibinfo {author} {\bibfnamefont
  {N.}~\bibnamefont {Marzari}},\ }\href
  {https://doi.org/https://doi.org/10.1016/j.cpc.2014.05.003} {\bibfield
  {journal} {\bibinfo  {journal} {Computer Physics Communications}\ }\textbf
  {\bibinfo {volume} {185}},\ \bibinfo {pages} {2309} (\bibinfo {year}
  {2014})}\BibitemShut {NoStop}%
\bibitem [{\citenamefont {Marzari}\ \emph {et~al.}(2012)\citenamefont
  {Marzari}, \citenamefont {Mostofi}, \citenamefont {Yates}, \citenamefont
  {Souza},\ and\ \citenamefont {Vanderbilt}}]{Marzari_mod}%
  \BibitemOpen
  \bibfield  {author} {\bibinfo {author} {\bibfnamefont {N.}~\bibnamefont
  {Marzari}}, \bibinfo {author} {\bibfnamefont {A.~A.}\ \bibnamefont
  {Mostofi}}, \bibinfo {author} {\bibfnamefont {J.~R.}\ \bibnamefont {Yates}},
  \bibinfo {author} {\bibfnamefont {I.}~\bibnamefont {Souza}},\ and\ \bibinfo
  {author} {\bibfnamefont {D.}~\bibnamefont {Vanderbilt}},\ }\href
  {https://doi.org/10.1103/RevModPhys.84.1419} {\bibfield  {journal} {\bibinfo
  {journal} {Rev. Mod. Phys.}\ }\textbf {\bibinfo {volume} {84}},\ \bibinfo
  {pages} {1419} (\bibinfo {year} {2012})}\BibitemShut {NoStop}%
\bibitem [{\citenamefont {Aryasetiawan}\ \emph {et~al.}(2004)\citenamefont
  {Aryasetiawan}, \citenamefont {Imada}, \citenamefont {Georges}, \citenamefont
  {Kotliar}, \citenamefont {Biermann},\ and\ \citenamefont
  {Lichtenstein}}]{crpa}%
  \BibitemOpen
  \bibfield  {author} {\bibinfo {author} {\bibfnamefont {F.}~\bibnamefont
  {Aryasetiawan}}, \bibinfo {author} {\bibfnamefont {M.}~\bibnamefont {Imada}},
  \bibinfo {author} {\bibfnamefont {A.}~\bibnamefont {Georges}}, \bibinfo
  {author} {\bibfnamefont {G.}~\bibnamefont {Kotliar}}, \bibinfo {author}
  {\bibfnamefont {S.}~\bibnamefont {Biermann}},\ and\ \bibinfo {author}
  {\bibfnamefont {A.~I.}\ \bibnamefont {Lichtenstein}},\ }\href
  {https://doi.org/10.1103/PhysRevB.70.195104} {\bibfield  {journal} {\bibinfo
  {journal} {Phys. Rev. B}\ }\textbf {\bibinfo {volume} {70}},\ \bibinfo
  {pages} {195104} (\bibinfo {year} {2004})}\BibitemShut {NoStop}%
\bibitem [{\citenamefont {Momma}\ and\ \citenamefont {Izumi}(2008)}]{VESTA}%
  \BibitemOpen
  \bibfield  {author} {\bibinfo {author} {\bibfnamefont {K.}~\bibnamefont
  {Momma}}\ and\ \bibinfo {author} {\bibfnamefont {F.}~\bibnamefont {Izumi}},\
  }\href {https://doi.org/10.1107/S0021889808012016} {\bibfield  {journal}
  {\bibinfo  {journal} {Journal of Applied Crystallography}\ }\textbf {\bibinfo
  {volume} {41}},\ \bibinfo {pages} {653} (\bibinfo {year} {2008})}\BibitemShut
  {NoStop}%
\end{thebibliography}%


\begin{thebibliography}{5}%
\makeatletter
\providecommand \@ifxundefined [1]{%
 \@ifx{#1\undefined}
}%
\providecommand \@ifnum [1]{%
 \ifnum #1\expandafter \@firstoftwo
 \else \expandafter \@secondoftwo
 \fi
}%
\providecommand \@ifx [1]{%
 \ifx #1\expandafter \@firstoftwo
 \else \expandafter \@secondoftwo
 \fi
}%
\providecommand \natexlab [1]{#1}%
\providecommand \enquote  [1]{``#1''}%
\providecommand \bibnamefont  [1]{#1}%
\providecommand \bibfnamefont [1]{#1}%
\providecommand \citenamefont [1]{#1}%
\providecommand \href@noop [0]{\@secondoftwo}%
\providecommand \href [0]{\begingroup \@sanitize@url \@href}%
\providecommand \@href[1]{\@@startlink{#1}\@@href}%
\providecommand \@@href[1]{\endgroup#1\@@endlink}%
\providecommand \@sanitize@url [0]{\catcode `\\12\catcode `\$12\catcode
  `\&12\catcode `\#12\catcode `\^12\catcode `\_12\catcode `\%12\relax}%
\providecommand \@@startlink[1]{}%
\providecommand \@@endlink[0]{}%
\providecommand \url  [0]{\begingroup\@sanitize@url \@url }%
\providecommand \@url [1]{\endgroup\@href {#1}{\urlprefix }}%
\providecommand \urlprefix  [0]{URL }%
\providecommand \Eprint [0]{\href }%
\providecommand \doibase [0]{https://doi.org/}%
\providecommand \selectlanguage [0]{\@gobble}%
\providecommand \bibinfo  [0]{\@secondoftwo}%
\providecommand \bibfield  [0]{\@secondoftwo}%
\providecommand \translation [1]{[#1]}%
\providecommand \BibitemOpen [0]{}%
\providecommand \bibitemStop [0]{}%
\providecommand \bibitemNoStop [0]{.\EOS\space}%
\providecommand \EOS [0]{\spacefactor3000\relax}%
\providecommand \BibitemShut  [1]{\csname bibitem#1\endcsname}%
\let\auto@bib@innerbib\@empty
\bibitem [{\citenamefont {Liu}\ \emph {et~al.}(2023)\citenamefont {Liu},
  \citenamefont {Araki}, \citenamefont {Arima}, \citenamefont {Itoh},
  \citenamefont {Asai},\ and\ \citenamefont {Masuda}}]{PRB_NISO}%
  \BibitemOpen
  \bibfield  {author} {\bibinfo {author} {\bibfnamefont {Z.}~\bibnamefont
  {Liu}}, \bibinfo {author} {\bibfnamefont {Y.}~\bibnamefont {Araki}}, \bibinfo
  {author} {\bibfnamefont {T.-h.}\ \bibnamefont {Arima}}, \bibinfo {author}
  {\bibfnamefont {S.}~\bibnamefont {Itoh}}, \bibinfo {author} {\bibfnamefont
  {S.}~\bibnamefont {Asai}},\ and\ \bibinfo {author} {\bibfnamefont
  {T.}~\bibnamefont {Masuda}},\ }\href
  {https://doi.org/10.1103/PhysRevB.107.064428} {\bibfield  {journal} {\bibinfo
   {journal} {Phys. Rev. B}\ }\textbf {\bibinfo {volume} {107}},\ \bibinfo
  {pages} {064428} (\bibinfo {year} {2023})}\BibitemShut {NoStop}%
\bibitem [{\citenamefont {Liechtenstein}\ \emph {et~al.}(1987)\citenamefont
  {Liechtenstein}, \citenamefont {Katsnelson}, \citenamefont {Antropov},\ and\
  \citenamefont {Gubanov}}]{Lichtenstein}%
  \BibitemOpen
  \bibfield  {author} {\bibinfo {author} {\bibfnamefont {A.}~\bibnamefont
  {Liechtenstein}}, \bibinfo {author} {\bibfnamefont {M.}~\bibnamefont
  {Katsnelson}}, \bibinfo {author} {\bibfnamefont {V.}~\bibnamefont
  {Antropov}},\ and\ \bibinfo {author} {\bibfnamefont {V.}~\bibnamefont
  {Gubanov}},\ }\href
  {https://doi.org/https://doi.org/10.1016/0304-8853(87)90721-9} {\bibfield
  {journal} {\bibinfo  {journal} {Journal of Magnetism and Magnetic Materials}\
  }\textbf {\bibinfo {volume} {67}},\ \bibinfo {pages} {65} (\bibinfo {year}
  {1987})}\BibitemShut {NoStop}%
\bibitem [{\citenamefont {Mazurenko}\ and\ \citenamefont
  {Anisimov}(2005)}]{Mazurenko_GF}%
  \BibitemOpen
  \bibfield  {author} {\bibinfo {author} {\bibfnamefont {V.~V.}\ \bibnamefont
  {Mazurenko}}\ and\ \bibinfo {author} {\bibfnamefont {V.~I.}\ \bibnamefont
  {Anisimov}},\ }\href {https://doi.org/10.1103/PhysRevB.71.184434} {\bibfield
  {journal} {\bibinfo  {journal} {Phys. Rev. B}\ }\textbf {\bibinfo {volume}
  {71}},\ \bibinfo {pages} {184434} (\bibinfo {year} {2005})}\BibitemShut
  {NoStop}%
\bibitem [{\citenamefont {He}\ \emph {et~al.}(2021)\citenamefont {He},
  \citenamefont {Helbig}, \citenamefont {Verstraete},\ and\ \citenamefont
  {Bousquet}}]{TB2J}%
  \BibitemOpen
  \bibfield  {author} {\bibinfo {author} {\bibfnamefont {X.}~\bibnamefont
  {He}}, \bibinfo {author} {\bibfnamefont {N.}~\bibnamefont {Helbig}}, \bibinfo
  {author} {\bibfnamefont {M.~J.}\ \bibnamefont {Verstraete}},\ and\ \bibinfo
  {author} {\bibfnamefont {E.}~\bibnamefont {Bousquet}},\ }\href
  {https://doi.org/https://doi.org/10.1016/j.cpc.2021.107938} {\bibfield
  {journal} {\bibinfo  {journal} {Computer Physics Communications}\ }\textbf
  {\bibinfo {volume} {264}},\ \bibinfo {pages} {107938} (\bibinfo {year}
  {2021})}\BibitemShut {NoStop}%
\bibitem [{\citenamefont {Ivanov}\ \emph {et~al.}(2013)\citenamefont {Ivanov},
  \citenamefont {Mathieu}, \citenamefont {Nordblad}, \citenamefont {Tellgren},
  \citenamefont {Ritter}, \citenamefont {Politova}, \citenamefont {Kaleva},
  \citenamefont {Mosunov}, \citenamefont {Stefanovich},\ and\ \citenamefont
  {Weil}}]{NISO_chem}%
  \BibitemOpen
  \bibfield  {author} {\bibinfo {author} {\bibfnamefont {S.~A.}\ \bibnamefont
  {Ivanov}}, \bibinfo {author} {\bibfnamefont {R.}~\bibnamefont {Mathieu}},
  \bibinfo {author} {\bibfnamefont {P.}~\bibnamefont {Nordblad}}, \bibinfo
  {author} {\bibfnamefont {R.}~\bibnamefont {Tellgren}}, \bibinfo {author}
  {\bibfnamefont {C.}~\bibnamefont {Ritter}}, \bibinfo {author} {\bibfnamefont
  {E.}~\bibnamefont {Politova}}, \bibinfo {author} {\bibfnamefont
  {G.}~\bibnamefont {Kaleva}}, \bibinfo {author} {\bibfnamefont
  {A.}~\bibnamefont {Mosunov}}, \bibinfo {author} {\bibfnamefont
  {S.}~\bibnamefont {Stefanovich}},\ and\ \bibinfo {author} {\bibfnamefont
  {M.}~\bibnamefont {Weil}},\ }\href {https://doi.org/10.1021/cm304095s}
  {\bibfield  {journal} {\bibinfo  {journal} {Chemistry of Materials}\ }\textbf
  {\bibinfo {volume} {25}},\ \bibinfo {pages} {935} (\bibinfo {year}
  {2013})}\BibitemShut {NoStop}%
\end{thebibliography}%

\section{Figure legends}
Figure 1: \textbf{a} Hexagonal cell of NISO. \textbf{b} Ni ions in the hexagonal cell. Only the closest neighbors coupled with exchange constants $J_1$ and $J_2$ are shown with yellow and gray lines, respectively. \textbf{c} $J_1$ and $J_2$ bonds from side view and top view around a Ni ion. \textbf{d} Definition of DM vector parameters. This figure explicitly depicts parameters for $\alpha=1$ bond type. \textbf{e} Electronic structure of a rhombohedral unit cell of NISO around Fermi level calculated within GGA (solid black line) and 2-orbital model constructed by MLWF method (cyan dashed line). The inset shows a schematic of the k-path in the Brillouin zone for the rhombohedral unit cell.

Figure 2: \textbf{a} Definition of the spin-spiral parameters. \textbf{b} The components of $\delta q$ minimizing the energy as a function of the spin rotation plane orientation (given by a polar angle $\phi$, where the rotation plane normal is $\hat{n}^{\perp}=(-\sin\phi,\cos\phi,0)$). The energy is minimized by $\delta q\sim(\cos\phi, \sin\phi,0)$ which is perpendicular to $\hat{n}^{\perp}$, giving rise to a cycloidal spiral.  The horizontal axis indicates the rotation of the spin-spiral plane. \textbf{c} The cycloidal spiral ground state for the Hamiltonian, with magnetic exchange constants derived from our superexchange theory. Black lines connecting the sites and spins of neighbors are drawn as a guide.

Figure 3: \textbf{a} Magnetic ground state of NISO as obtained from classical Monte-Carlo simulations. The red rectangle indicates the area, for which we show the spin texture in Panel \textbf{b}. \textbf{b} Magnetic structure inside the area, indicated by the red box in Panel \textbf{a} for different values of an external magnetic field, perpendicular to the spiral plane (three-fold rotation axis).

Figure 4: \textbf{a}-\textbf{d} Continuous spin, \textbf{e}-\textbf{h} real space continuous spin textures and \textbf{i}-\textbf{l} sphere area covered by each spin array as calculated using the continuous model Eq.(\ref{eqn:Enrec}) at several external magnetic field strengths $H_z$. \textbf{Panels a, e, i:} SF state; \textbf{panels b, f, j:} conical spiral; \textbf{panels c, g, k:} kink state; \textbf{panels d, h, l:} flat spiral state.

Figure 5: The response to an external magnetic field $H_z$, applied along the three-fold axis in NISO. (Upper panel) Spin spiral period calculated from the continuous model (red line) and using the analytical formula in Ref.~\cite{DeGennes} (blue line). (Lower panel) magnetically-induced polarization resulting from the isotropic exchange striction. The dotted line indicates phase transition between the flat spiral state and the conical state shown in Fig.~6.

Figure 6: Phase diagram of NISO as obtained by minimization of the continuous model, Eq.(\ref{eqn:Enrec}), with a small electric field  $E_z$. Yellow, green, and red areas are indicating flat spiral phase, conical phase, and SF phase, respectively. The tendency towards FM state is indicated with white in color gradients.

Figure 7: \textbf{a} Schematic representation of the SF phase. \textbf{b} Evolution of the angle $\theta$ after SF transition under the external magnetic field as obtained from the MCS. \textbf{c} Polarization change during the transition from SF phase to FM state.


\end{document}


\title{Supplementary Information: Multiferroic kinks and spin-flop transition in Ni$_{2}$InSbO$_6$ from first principles}

\author{Ryota Ono}
    \email[Correspondence email address: ]{ryota.ono.gm@gmail.com} 
    \affiliation{Quantum Materials Theory, Italian Institute of Technology, Via Morego 30, 16163 Genova, Italy}
    \affiliation{Research Center for Materials Nanoarchitectonics (MANA), National Institute for Materials Science (NIMS), 1-1 Namiki, Tsukuba, Ibaraki 305-0044, Japan}
\author{Igor Solovyev}
    \affiliation{Research Center for Materials Nanoarchitectonics (MANA), National Institute for Materials Science (NIMS), 1-1 Namiki, Tsukuba, Ibaraki 305-0044, Japan}
\author{Sergey Artyukhin}
    \affiliation{Quantum Materials Theory, Italian Institute of Technology, Via Morego 30, 16163 Genova, Italy}


\maketitle

\section{Supplementary note 1: Symmetric anisotropy tensors and its effects on magnetism}
In this section, we present the values of symmetric anisotropy tensors derived from our analytical formulas and resulting effects on the magnetic ground state in NISO. 
These tensors are subject to symmetry constraints imposed by the $\hat{C}_3^z$ rotational symmetry. 
For example, $\Gamma_{01}^{xz}=-\frac{1}{2} \Gamma_{02}^{xz}+\frac{\sqrt{3}}{2}\Gamma_{02}^{yz}=-\frac{1}{2} \Gamma_{03}^{xz}-\frac{\sqrt{3}}{2}\Gamma_{03}^{yz}$. 
This symmetry property can be confirmed by operating $\hat{C}_3^z$ on spin at the origin (center of the rotation) $S_{0}$ and the bond $S_{j}$ (upper layer: $j=1,2,3$, lower layer: $j=4,5,6$), and compare the energy equivalence.

\begin{figure}[htbp]
\centering
\includegraphics[keepaspectratio, scale=0.6]{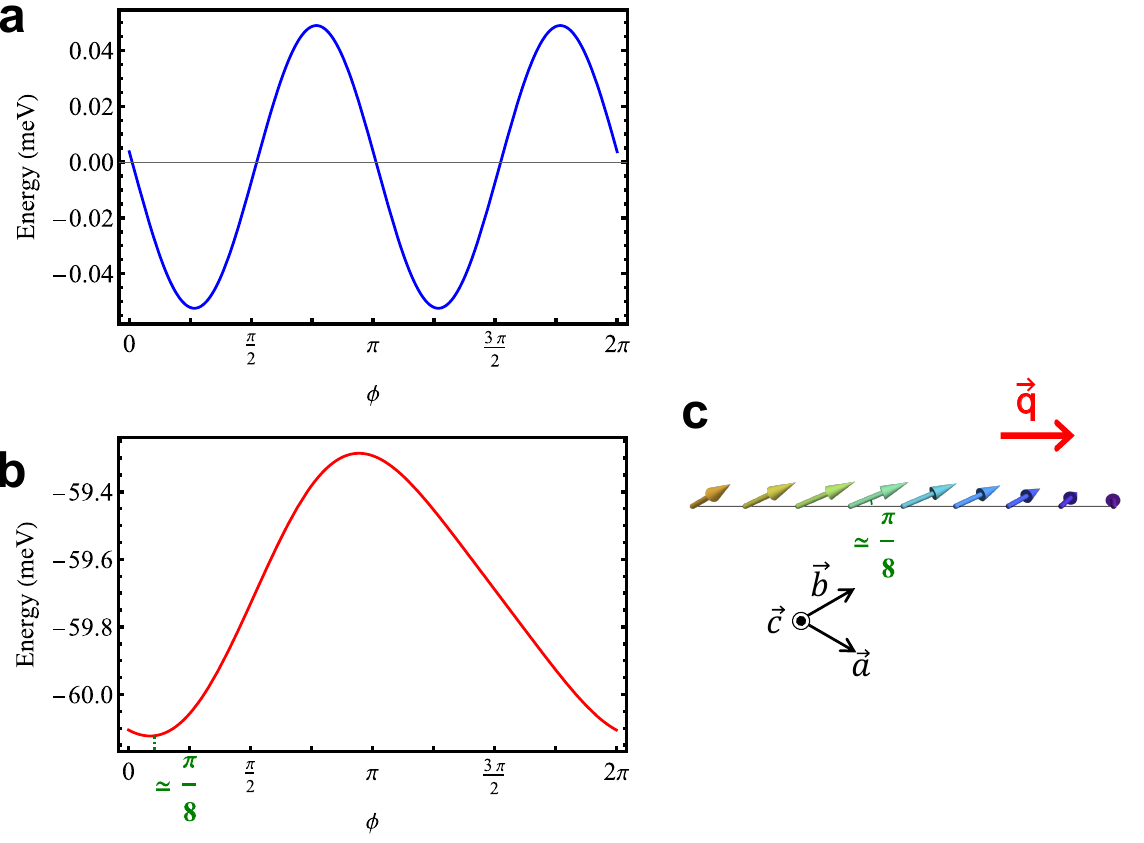}
\caption{Spin-spiral plane rotation induced by the symmetric anisotropic exchange interactions. \textbf{a} The energy change by the spiral plane rotation solely due to the symetric anisotropic exchange interactions. \textbf{b} Total energy change by the spiral plane trotation. \textbf{c} A schematic for the spin-spiral state given as the minima of \textbf{b}. }\label{fig:E_sym}
\end{figure}

The symmetric anisotropic exchanges is another important factor in determining spin-spiral type. 
We discuss such effects here.
First, we fix spin-spiral propagation vector as $\vec{q} = (\delta q, 0, 0)$. 
Then, using Eq.~(6) of the main text, we can evaluate energy change by the rotation of $\phi$.
In this setting, $\phi=n\pi$ and $\phi=n\pi/2$ $(n=0,1,\dots)$ correspond to proper-screw and cycloidal, respectively.
The obtained energy change is shown in Fig.~\ref{fig:E_sym} \textbf{a}.
The results indicate energy minima is shifted to $\phi \approx \pi/4$.
Thus, the symmetric anisotropy prefers spin-spiral rotation plane to be between proper-screw and cycloidal.  
On the other hand, DM interactions prefer $\phi=0$.
The total energy change is shown in Fig.~\ref{fig:E_sym} \textbf{b}.
The minima is shifted to $\phi \approx \pi /8$.
Thus, we expect symmetric anisotropic exchange makes spiral to be slightly rotated from a perfect cycloidal as shown in Fig.~\ref{fig:E_sym} \textbf{c}.

\begin{table}
\caption{Values of symmetric anisotropic exchanges for bonds given in Fig.~1\textbf{c} [meV]} 
\label{tab:anisotropy}
\begin{ruledtabular}
\begin{tabular}{ccc}
 & $\alpha=1, (j=1,2,3): $& \\
$\stackrel{\leftrightarrow}{\Gamma}_{01}= 
\begin{pmatrix}
 0.062 & -0.023 & 0.021 \\
 -0.023 & -0.052 & 0.100 \\
 0.021 & 0.100 & -0.010 \\
\end{pmatrix}$ &
$\stackrel{\leftrightarrow}{\Gamma}_{02} = 
\begin{pmatrix}
 -0.052 & -0.032 & -0.102 \\
 -0.032 & 0.060 & -0.024 \\
 -0.102 & -0.024 & -0.008 \\
\end{pmatrix}$ &
$\stackrel{\leftrightarrow}{\Gamma}_{03} = 
\begin{pmatrix}
 -0.000 & 0.063 & 0.071 \\
 0.063 & 0.006 & -0.071 \\
 0.071 & -0.071 & -0.006 \\
\end{pmatrix}$ \\
\\
 & $\alpha=2, (j=4,5,6): $& \\
 $\stackrel{\leftrightarrow}{\Gamma}_{04} = 
\begin{pmatrix}
 0.033 & -0.056 & 0.041 \\
 -0.056 & -0.050 & 0.070 \\
 0.041 & 0.070 & 0.017 \\
\end{pmatrix}$ &
$\stackrel{\leftrightarrow}{\Gamma}_{05} = 
\begin{pmatrix}
 0.024 & 0.061 & 0.044 \\
 0.061 & -0.035 & -0.074 \\
 0.044 & -0.074 & 0.011 \\
\end{pmatrix}$ &
$\stackrel{\leftrightarrow}{\Gamma}_{06} = 
\begin{pmatrix}
 -0.089 & -0.005 & -0.092 \\
 -0.005 & 0.070 & 0.003 \\
 -0.092 & 0.003 & 0.019 \\
\end{pmatrix}$ \\
\end{tabular}
\end{ruledtabular}
\end{table}

\section{Supplementary note 2: Single-ion anisotropy energy from 2-orbital model}
\begin{figure}[htbp]
\centering
\includegraphics[keepaspectratio, scale=0.6]{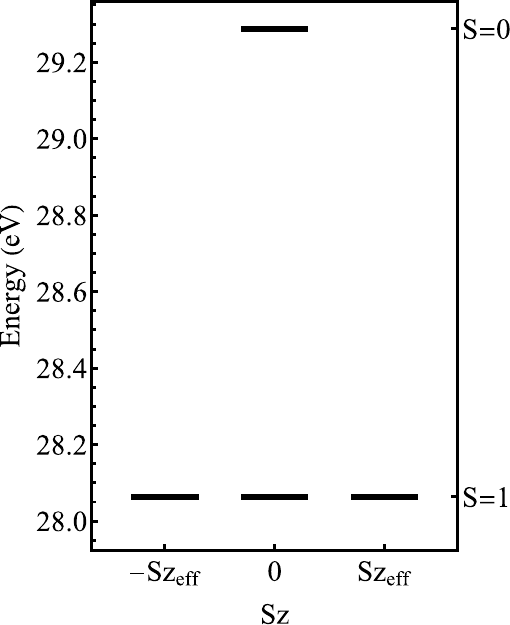}
\caption{Energies of two-electron state on a Ni 2-orbital model as obtained from diagonalization of the on-site Hamiltonian.}\label{fig:Elevel}
\end{figure}

To confirm the magnitude of the single-ion anisotropy (SIA) in our model, we perform a diagonalization of the on-site Hamiltonian, given as $\mathcal{H_{\mathrm{on-site}}} = \mathcal{H_{\mathrm{CF}}} + \mathcal{H_{\mathrm{SOC}}} + \mathcal{H_{\mathrm{U}}}$, where $\mathcal{H_{\mathrm{CF}}}$, $\mathcal{H_{\mathrm{SOC}}}$ and $\mathcal{H_{\mathrm{U}}}$ indicate crystal-field, SOC, and on-site Coulomb interaction, respectively.
Obtained energy spectrum is depicted in Fig.~\ref{fig:Elevel}.
After the diagonalization, the $Sz$ values are reduced to $Sz_{\mathrm{eff}}\approx 0.7$ due to SOC. 
The SIA is determined by the energy difference within the $S=1$ multiplet.
It is important to note that, due to the influence of SIA, these states do not strictly form a triplet.
The energy difference between doubly degenerate $\pm Sz_{\mathrm{eff}}$ states and non-degenerate $Sz=0$ state is  $E_{\pm Sz_{\mathrm{eff}}} - E_{Sz=0} \approx 7$~$\mu$eV. 
Indicating very tiny easy-plane anisotropy, that is consistent with the expectation from the experiment~\cite{PRB_NISO}.
Consequently, in our 2-orbital model, the SIA can be considered negligible.

\section{Supplementary note 3: Anisotropic mechanism for magnetically induced polarization}
Here, we discuss the effect of the anisotropic mechanism for the magnetically-induced polarization (the second and the third term in Eq.~(3) of the main text) in NISO.
We summerize the parameters for the antisymmetric mechanism of the magnetically-induced polarization in Table~\ref{tab:antis}.
In all bonds, the z-component (the third row of the tensor) of this term results in small compared to the in-plane component (the first and the second row of the tensor).
When we take only z-component and write it as $\vec{\mathbf{\mathcal{P}}}_{0j,z}^\alpha$ for each bond type $\alpha$, this vector transform between the same $\alpha$ exactly same as the DM vector (without unphysical numerical error due to position operator).
Namely, using the same definition of DM vector (Eq.~(4) of the main text), one can write it as $\vec{\mathbf{\mathcal{P}}}_{0j,z}^\alpha = \left( \mathbf{\mathsf{p}}_{\alpha,z}^{\parallel} \cos\theta'_{j,\alpha}, \mathbf{\mathsf{p}}_{\alpha,z}^{\parallel} \sin\theta'_{j,\alpha}, \mathbf{\mathsf{p}}_{\alpha,z}^{\perp} \right)$.
Now, we take into account the long-wave length spin-spiral ground state propagate in the Ni plane as obtained from our analysis and the prior experimental observation.
Then, neighboring sites for the same $\alpha$ can be given as 
Since the spiral rotates in a plane formed by $(0,0,1)$ and an arbitrary vector in $xy$-plane $(\cos\phi, \sin\phi, 0)$, the total polarization along $z$-direction will vanish.
On the other hand, the $xy$ components of the polarization can be finite.
Using the values of antisymmetric mechanism given in Table~\ref{tab:antis}, regardless of the spiral type (cycloidal or proper-screw), $P_x$ and $P_y$ are evaluated as $P_x \approx -16 $~$\mu C /m^2$ and $P_y \approx -9$~$\mu C /m^2$, respectively.
Therefore, polarization from the antisymmetric term can be neglected in the main discussion.

The parameters for the symmetric anisotropic term of the magnetically-induced polarization are shown in Table~\ref{tab:syman}.
The symmetric anisotropic term for the magnetically-induced polarization results in very tiny. 
This is expected from the formula Eq.~(7) of the main text.
The symmetric aniostropy term of the polarization originates from spin off-diagonal term of hoppings and position operators.
As one can expect from the relation between isotropic and antisymmetric parameters, we obtain few ~$\mu C /m^2$ order for the symmetric anisotropic term.
Thus, we can safely ignore this term in the main discussions.

\begin{table}
\caption{Values of antisymmetric mechanism of the magnetically-induced polarization for bonds given in Fig.~1\textbf{c} in the main text [$\mu C /m^2$]} 
\label{tab:antis}
\begin{ruledtabular}
\begin{tabular}{ccc}
 & $\alpha=1, (j=1,2,3): $& \\
$\vec{\mathbf{\mathcal{P}}}_{01}= 
\begin{pmatrix}
 1.26 & -3.37 & 1.36 \\
 -57.33 & 29.93 & 27.12 \\
 2.52 & 0.28 & -6.61 \\
\end{pmatrix}$ &
$\vec{\mathbf{\mathcal{P}}}_{02} = 
\begin{pmatrix}
 48.09 & 29.30 & 22.41 \\
 -26.06 & -16.29 & -16.65 \\
 -1.97 & -0.86 & -8.14 \\
\end{pmatrix}$ &
$\vec{\mathbf{\mathcal{P}}}_{03} = 
\begin{pmatrix}
 -8.14 & 54.98 & -26.74 \\
 -1.87 & 35.12 & -13.64 \\
 -2.01 & 0.92 & -6.46 \\
\end{pmatrix}$ \\
\\
 & $\alpha=2, (j=4,5,6): $& \\
 $\vec{\mathbf{\mathcal{P}}}_{04} = 
\begin{pmatrix}
 69.59 & 12.95 & 14.59 \\
 44.57 & -7.87 & 14.78 \\
 -7.18 & 16.92 & -7.73 \\
\end{pmatrix}$ &
$\vec{\mathbf{\mathcal{P}}}_{05} = 
\begin{pmatrix}
 -11.65 & 3.94 & 3.88 \\
 30.17 & 69.00 & -16.21 \\
 15.56 & -5.08 & -5.14 \\
\end{pmatrix}$ &
$\vec{\mathbf{\mathcal{P}}}_{06} = 
\begin{pmatrix}
 34.12 & -55.18 & -18.70 \\
 -30.29 & 22.44 & 4.21 \\
 -12.68 & -10.86 & -7.19 \\
\end{pmatrix}$ \\
\end{tabular}
\end{ruledtabular}
\end{table}

\begin{table}
\caption{Values of the symmetric anisotropic mechanism of the magnetically-induced polarization for bonds given in Fig.~1\textbf{c} [$\mu C /m^2$]} 
\label{tab:syman}
\begin{ruledtabular}
\begin{tabular}{ccc}
 & $\alpha=1, (j=1,2,3): $& \\
$\stackrel{\leftrightarrow}{\Pi}_{01}^z =
\begin{pmatrix}
      -0.196   &   -2.110   &   -2.224 \\
      -2.110   &   0.454   &    2.981 \\
      -2.224   &    2.981   &   -0.257 \\
\end{pmatrix}$ &
$\stackrel{\leftrightarrow}{\Pi}_{02}^z = 
\begin{pmatrix}
       2.086   &    0.792    &   3.651 \\
       0.792   &   -1.760    &   0.421 \\
       3.651   &    0.421    &  -0.325 \\
\end{pmatrix}$ &
$\stackrel{\leftrightarrow}{\Pi}_{03}^z = 
\begin{pmatrix}
      -1.633   &    1.284   &   -1.538 \\
       1.284   &    1.822   &   -3.510 \\
      -1.538   &   -3.510   &   -0.188 \\
\end{pmatrix}$ \\
\\
 & $\alpha=2, (j=4,5,6): $& \\
 $\stackrel{\leftrightarrow}{\Pi}_{04}^z = 
\begin{pmatrix}
       1.256   &    1.130   &    2.070 \\
       1.130   &   -0.978   &    0.373 \\
       2.070   &    0.373   &   -0.277 \\
\end{pmatrix}$ &
$\stackrel{\leftrightarrow}{\Pi}_{05}^z = 
\begin{pmatrix}
      -1.363   &    0.457   &   -0.745 \\
       0.457   &    1.674   &   -2.062 \\
      -0.745   &   -2.062  &   -0.310 \\
\end{pmatrix}$ &
$\stackrel{\leftrightarrow}{\Pi}_{06}^z = 
\begin{pmatrix}
       0.442   &   -1.556   &   -1.469 \\
      -1.556   &   -0.167   &    1.648 \\
      -1.469   &    1.648   &   -0.274 \\
\end{pmatrix}$ \\
\end{tabular}
\end{ruledtabular}
\end{table}

\section{Supplementary note 4: Green's function method}
The exchange parameters can be calculated from DFT+U method.
There are a few methods available to calculate them from DFT+U. 
One of the most widely adopted techniques is the Green’s function method~\cite{Lichtenstein}. 
The method starts from the local force theorem, that connects small perturbation energy from the ground state and sum of one-particle energy changes for the occupied states.

By employing Lloyd's formula, it is possible to represent changes in the electron density of states resulting from perturbations to the Hamiltonian and Green's function.
Then, the idea is to establish a connection between this perturbation energy and the energy change in the spin Hamiltonian due to infinitesimal rotations of spins. 
As a result, for example, isotropic exchange is evaluated as:
\begin{align}
    J_{ij} = \frac{1}{4\pi} \int^{E_F}_{-\infty} d\epsilon \; \mathrm{Im} \; \mathrm{Tr}_{m} \left( \Delta_i G_{ij}^{\downarrow} \Delta_j G_{ji}^{\uparrow} \right),
\end{align}
where the trace runs over the magnetic quantum number $m$, $\Delta_i$ represents the magnetic splitting at site $i$, and $G_{ij}$ denotes the Green's function.
The expansion of this method to include SOC can be derived using the Dyson equation of the Green's function, as discussed in Ref.~\cite{Mazurenko_GF}. 
We use this Green's fucntion method as it is implemented in the TB2J package~\cite{TB2J}.
The calculated exchange parameters from a layer AFM state are given in Tab.~\ref{tab:Green}.

\begin{figure}[htbp]
\centering
\includegraphics[keepaspectratio, scale=1]{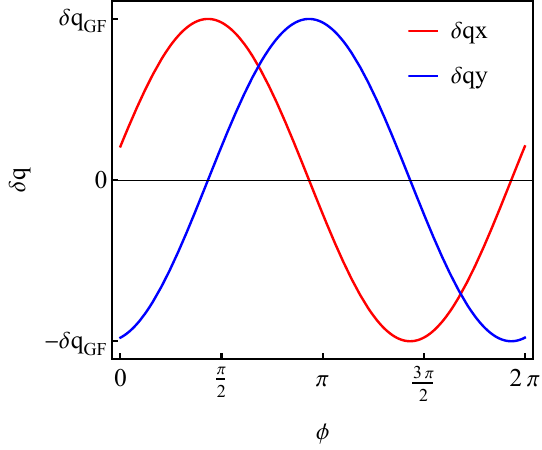}
\caption{Single-$q$ Ansatz result for the exchange parameters calculated using Green's function method.}\label{fig:spiral}
\end{figure}

The anisotropic exchange parameters yield magnitudes quite different from those obtained by our analytical theory.
This suppression of anisotropic exchanges leads to a long-wavelength solution of the spin-spiral Ansatz. 
By employing the Ansatz equations Eq.~(9) of the main text, we find the ground state to be approximately proper-screw state (Fig.~\ref{fig:spiral}).
The wavelength results in a very long $\delta q_{\mathrm{GF}} \approx 0.007$ (equivalent to 142 unit cells) due to the above-mentioned suppression of the anisotropic exchange.

\begin{table}
\captionof{table}{Values of magnetic exchanges in NISO [$meV$] and corresponding parameters calculated by Green's function method }\label{tab:Green}
\begin{ruledtabular}
\begin{tabular}{cccccc}
Bond $\alpha$ & Dist. & $J_\alpha$ & $d^{\parallel}_\alpha$ & $d^{\perp}_\alpha$ & $\theta_\alpha$~[Deg.]\\
\hline
1 & 3.821 & -6.524 & 0.651 &  0.022 & 97 \\
2 & 3.912 & -12.275 & 0.668 & 0.090 & -125  \\  
\end{tabular}
\end{ruledtabular}
\end{table}

\section{Supplementary note 5: Critical temperature by Monte-Carlo simulation}
\begin{figure}[htbp]
\centering
\includegraphics[keepaspectratio, scale=0.6]{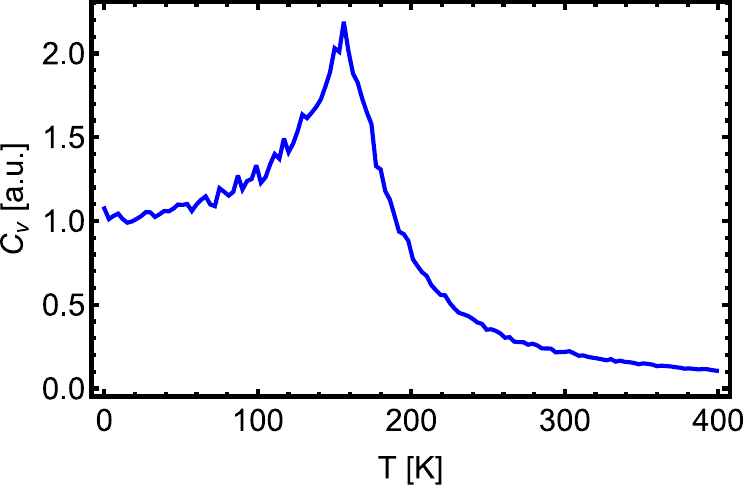}
\caption{Heat capacity as obtained from Monte-Carlo simulation}\label{fig:Cv}
\end{figure}

To investigate the transition temperature, we calculate the heat capacity. 
The heat capacity calculated from the Monte-Carlo simulations is shown in Fig.~\ref{fig:Cv}.
The result indicates the phase transition around 150~K. 
This is exaggerated compared to the experimental value 77~K~\cite{NISO_chem}.
This inconsistency arises from finite size dependence and the overestimated exchange parameters in the 2-orbital model.
Given the relationship between the calculated transition temperature and the experimental value, the exchange parameters appear to be overestimated by approximately a factor of two.

\bibliography{ref_sup.bib}
